\documentclass[sn-basic,iicol, Numbered]{sn-jnl}

\usepackage{amsmath,amssymb, amsthm}
\usepackage{graphicx}
\usepackage{booktabs}
\usepackage{tabularx}
\usepackage{bm}
\usepackage{algorithm} 
\usepackage{algpseudocode}   
\usepackage{standalone}
\usepackage{tikz}
\usepackage{comment}

\newcommand{\R}{\mathbb{R}}
\newcommand{\E}{\mathbb{E}}

\newcommand{\bx}{\bm{x}}
\newcommand{\bt}{\bm{t}}

\newcommand{\bv}{\bm{v}}
\newcommand{\bz}{\bm{z}}

\newcommand{\bR}{\bm{R}}

\newcommand{\bu}{\bm{u}}
\newcommand{\bU}{\mathbf{U}}

\newcommand{\btheta}{\bm{\theta}}

\newcommand{\bG}{\mathbf{G}}

\newcommand{\bK}{\mathbf{K}}
\newcommand{\bH}{\bm{H}}

\newcommand{\bS}{\bm{S}}
\newcommand{\bP}{\bm{P}}

\newcommand{\bbeta}{\bm{\beta}}
\newcommand{\balpha}{\bm{\alpha}}
\newcommand{\bmu}{\bm{\mu}}
\newcommand{\bgamma}{\bm{\gamma}}
\newcommand{\bGamma}{\bm{\Gamma}}
\newcommand{\bpi}{\bm{\pi}}
\newcommand{\bphi}{\bm{\phi}}
\newcommand{\bPhi}{\bm{\Phi}}

\newcommand{\eye}{\mathbf{I}}

\newcommand{\N}{\mathcal{N}}

\newcommand{\K}{\mathcal{K}}
\newcommand{\LL}{\mathcal{L}}
\newcommand{\HH}{\mathcal{H}}
\newcommand{\A}{\mathcal{A}}

\newtheorem{lemma}{Lemma}
\newtheorem{proposition}{Proposition}
\newtheorem*{remark}{Remark}
\newtheorem{definition}{Definition}

\DeclareMathOperator{\diag}{diag}
\DeclareMathOperator{\peq}{\stackrel{+}{=}}

\begin{document}

\title{Koopman Model Dimension Reduction via Variational Bayesian Inference and Graph Search } %

\author*[1]{\fnm{Selin Ezgi} \sur{Ozcan}}\email{seozcan@tudelft.nl}

\author[1]{\fnm{Mustafa Mert} \sur{Ankaralı}}\email{mertan@metu.edu.tr}

\affil*[1]{\orgdiv{Electrical and Electronics Engineering}, \orgname{Middle East Technical University}, \orgaddress{\city{Ankara}, \country{Turkiye}}}


\date{\today}

\abstract{Koopman operator recently gained increasing attention in the control systems community for its abilities to bridge linear and nonlinear systems.
Data driven Koopman operator approximations have established themselves as key enablers for system identification and model predictive control.
Nonetheless, such methods commonly entail a preselected definition of states in the function space
leading to high dimensional, overparameterized models that may suffer from poor numerical conditioning or degraded long term prediction performance.
We address the preselected states problem by proposing a hierarchical probabilistic approach for Koopman model identification to enable dictionary component selection.
In our method, elements of the model are treated as random variables and the posterior estimates are found using variational Bayesian (VB) inference updates.
Posterior inclusion probabilities define a directed graph of regressor–target dependencies. Graph ancestry search then selects the observables required by prescribed outputs in the thresholded model.
We demonstrate that the proposed reduction can improve numerical conditioning and can preserve or improve prediction performance while substantially reducing the dictionary size.

}

\maketitle %

\section{Introduction}\label{sec:intro}
In recent years, Koopman operator has attracted significant attention in various fields of control systems due to its ability to represent nonlinear dynamical systems with globally linear models \cite{shi2024koopmansurvey}.
Rather than linearizing the system around specific operating points in the state space, the Koopman approach introduces exact linear models in the function space \cite{koopman1931hamiltonian}.
The dynamics in the function space is linear, so the operator enables the use of a vast amount of well established linear systems tools on nonlinear systems, significantly improving the workflow.
Despite its promise, the Koopman operator is inherently infinite dimensional and not directly applicable in real world settings.
Therefore, finite dimensional approximations of the operator are needed for practical implementation \cite{mauroy2020koopman}.

Pioneering works have fostered the adoption of the data driven finite dimensional approximation of the Koopman operator particularly in system identification and model based control \cite{dmd, edmd, koopmanwithcontrol, abraham2017model}.
These methods approximate the operator on a finite dimensional subspace which is spanned by an a priori selected dictionary of functions.
It is common in the literature to denote the functions in the dictionary as observables.
Based on the values of the observables on the measured dataset, linear regression is performed to estimate the Koopman operator as a matrix \cite{dmd, edmd}.
The eigenspace of this matrix then provides detailed and interpretable information about the system behavior, which has proven to be valuable in a wide range of system identification and control applications \cite{mauroy2020koopman}.

Selecting a compact and informative observable dictionary remains an important design problem in finite dimensional Koopman modeling. 
Although systematic approaches have been developed, the appropriate selection criterion depends on the modeling objective, the candidate observables and the available data.
The a priori selected dictionary plays a major role in the description and prediction quality of the model \cite{pan2023tropaper, khosravi, lusch2018deep}.
If the dictionary adequately captures the dominant nonlinearities and yields numerically well conditioned optimization problems, the resulting Koopman model can exhibit strong generalization performance.
Yet, the selection of an appropriate dictionary is highly system dependent and no universal choice exists \cite{pan2023tropaper}. 
Some approaches rely on encoder-decoder architectures to implicitly learn the dictionary \cite{lusch2018deep}. 
Deep learning solutions \cite{koopmandictionary} usually can learn the embeddings well, but the deep models may not be as easily interpretable as standard approaches.
Neural dictionary learning can involve substantial computational and data requirements, depending on the architecture and application. This motivates considering simpler dictionary selection methods alongside learned representations \cite{thompson2022computationallimitsdeeplearning}.
This prioritizes more conventional methods whenever possible.
Traditional Koopman approximation methods commonly use predetermined polynomial bases, radial basis functions (RBF), particularly squared exponential kernels and Fourier features in the dictionary \cite{mauroy2020koopman}.
Even when a particular class of observables is selected for the dictionary, the parameters related to the functions remain unspecified. 
For example, the exponent coefficient of squared exponential kernels or polynomial degree must also be determined.
Moreover, the chosen observables may cause ill conditioned problems 
since observables are nonlinear functions of the measured states, measurement noise is transformed in a nontrivial and generally unknown manner. 
As a result, the noise characteristics of the observables cannot be quantified a priori and the nonlinear mapping may significantly amplify or distort the noise, thereby degrading numerical conditioning \cite{vb}.
Another open design question is the size of the dictionary.
In principle, the span of the dictionary must be invariant under the operator.
However, this is often not realizable in the finite dimensional setting for practical scenarios.
In the finite dimensional setting, this requirement corresponds to ensuring that the projection of the finite dimensional operator’s image onto the chosen dictionary incurs only a small approximation error \cite{mauroy2020koopman}.
Adding more observables does not necessarily reduce the approximation error, and may even exacerbate prediction errors \cite{pan2023tropaper}.

These open problems reveal the need for systematic assessment of the dictionary and informed observable selection.
\cite{brunton2016discovering} selects the regression variables by sequential thresholded least squares approach, where observables are pruned based on the magnitudes of their weights.
This approach treats each target variable independently, but does not by itself enforce dependency closure of a common observable dictionary.
Furthermore, it does not account for how differences in the scaling of observables affect the resulting weights.
Moreover, setting a single threshold for various types of observables is not very meaningful in the physical sense as each observable and state may have different physical units.
Another work \cite{pan2023tropaper} highlights the misleading performance of the one step prediction error metric as the objective for the Koopman model problems and demonstrates that some Koopman models can perform reasonably well in one step prediction but they actually fail to do prediction in long term, which is crucial for model predictive control.
They assign an inclusion variable to each dictionary element and set them as hyperparameters to a control related cost function, which they optimize using tree structured Parzen estimators \cite{TPE}.

Several principled approaches address the selection of informative Koopman representations. 
\cite{pan2021sparsity} develop sparsity promoting methods for EDMD and KDMD that select informative Koopman invariant subspaces by pruning redundant or spurious spectral components and assessing their departure from linear evolution. 
\cite{schlosser2022sparsity} establish how sparse dependencies in an underlying dynamical system induce subsystem decompositions and corresponding structures in the Koopman and Perron Frobenius operators. 
These contributions provide complementary foundations for selecting compact representations and exploiting dynamical structure.

Building on these ideas, we investigate dictionary screening for a prescribed set of output observables. 
Starting from a fixed candidate dictionary, we use a hierarchical Bernoulli Gaussian regression model and variational Bayesian inference to estimate inclusion probabilities for individual regressor target pairs. 
Thresholding these probabilities defines a directed dependency graph among the observables. Ancestry search then identifies the observables required by the selected outputs under the inferred dependency structure. 
The contribution is the integration of probabilistic coefficient inclusion with graph-based, output oriented dictionary screening. 
For the thresholded finite dimensional predictor, we establish output preservation under restriction to the retained coordinates; the selected dictionary can subsequently be used to refit the predictor.

Related work \cite{Nayek_2021} uses probabilistic priors and Gibbs sampling to select relevant basis functions for discovery of nonlinear governing equations. 
Consequently, \cite{Nayek_2021} provides a precedent for probabilistic inclusion based variable selection in dynamical system identification.
Our approach uses variational inference for a finite dimensional Koopman model and combines the inferred regressor-target inclusion probabilities with graph based ancestry search to select a dictionary for prescribed output observables. 

Our contributions in this paper are listed as:
\begin{itemize}
    \item We provide hierarchical probabilistic modeling of the Koopman problem that serves as the intermediate tool between initial dictionary proposal and the reduced dictionary which is supported by data.
    Our formulation distinguishes weight magnitudes from inclusion in the model.
    \item We derive the VB inference updates for our proposed model.
    \item By resorting to graph techniques, we reduce the Koopman model dimension.
    We establish output preservation for restriction of the thresholded finite dimensional predictor to the retained coordinates.
    \item We show that other traditional tools for Koopman model identification benefit from the reduced dictionary after our proposed algorithm is applied.
    \item We demonstrate the performance of our dictionary reduction method in both simulated and real world settings. 
    Specifically, we show that the proposed hierarchical model enables considerable dictionary reduction and it can improve or retain the performance of the models with the initial full size dictionary.
\end{itemize}

This paper is organized as follows. 
In Section \ref{sec:background}, we provide the preliminaries about the Koopman operator, VB inference and basic graph condensation. 
Section \ref{sec:PS} outlines the problem and introduces the probabilistic model which we propose.
Then in Section \ref{sec:method}, we derive the VB updates and provide our dictionary reduction algorithm based on the inclusion random variables.
In Section \ref{sec:results}, we demonstrate the results of our experiments.
Lastly, a conclusion and discussion is given in Section \ref{sec:conclusion}.

\section{Preliminaries}\label{sec:background}
\subsection{Koopman Operator and Data Driven Approximations}
Let a discrete time dynamical system be described by the map $\bS: X \to X$,
\begin{align}
    \bx[k+1] = \bS(\bx[k]),
\end{align}
where $X$ is the state space of the system and $k$ is the discrete time index.%
In our context, we simply assume $X \subset \R^d$.
An observable $\psi \in \HH$ is defined as a scalar valued function of the state $\psi: X \to \R$, where $\HH$ is a Hilbert space.
Letting $\LL(\HH)$ be the operator space on $\HH$, 
the discrete time Koopman operator $\K \in \LL(\HH)$ is defined through its action on the observables by the system specific composition \cite{koopman1931hamiltonian}
\begin{equation}
    \K \psi = \psi \circ \bS.
\end{equation}

The Koopman operator is linear in its arguments and thus provides a global linear representation for the system unlike linearization around a point in the state space.
However, the operator suffers from possible infinite dimensionality \cite{mauroy2020koopman}.

Data driven algorithms such as Extended Dynamic Mode Decomposition (EDMD) have been proposed to circumvent the infinite dimensionality problem \cite{edmd}. 
EDMD aims to approximate the operator in a finite dimensional subspace spanned by a predefined set of observables and express the action of the operator as a matrix with a predefined set of observables as the basis \cite{mauroy2020koopman, edmd}.
Let the dictionary of observables be \( \bphi(\bx) = \begin{bmatrix} \phi_1(\bx) \ \ldots \ \phi_L(\bx) \end{bmatrix}^T \) and $F \in \HH$ be the subspace spanned by $\bphi$.
Since $F$ is a finite dimensional subspace, the projected restriction of $\K$ from and onto $F$ can be represented as a matrix $\bK_F$ with the basis $\bphi$.
The aim of EDMD is to approximate $\bK_F$ as $\hat{\bK}_F$ through least squares solutions of the optimization problem \cite{mauroy2020koopman, edmd}:
\begin{equation}
    \hat{\bK}_F = \underset{\bK_F \in \R^{L \times L}}{\arg \min} \sum_{i=1}^M || \bphi(\bx[i+1])^T - \bphi(\bx[i])^T\bK_F ||^2.
\end{equation}
One of the most useful extensions of the algorithm adapts the approach for controlled systems while maintaining the basic workflow \cite{koopmanwithcontrol}.
At the end, EDMD with control input models approximate a linear state space structure in terms of the dictionary elements as the states:
\begin{align}
    \begin{split}
    \bphi(\bx[k+1])^T &= [\bphi(\bx[k])^T \bu[k]^T]\hat{\bK}_F \\
    \bx[k] &= \mathbf{C} \bphi(\bx[k]).        
    \end{split}
    \label{eq:koopman_model}
\end{align}

The Koopman matrix $\hat{\bK}_F$ can be decomposed into the state space matrices 
\begin{align}
    \hat{\bK}_F = \begin{bmatrix}
        \mathbf{A}^T \\ \mathbf{B}^T
    \end{bmatrix},
    \label{eq:KF_blocks}
\end{align}
which concludes the construction of the linear Koopman model.

\subsection{Variational Bayesian (VB) Inference}
The Koopman EDMD model identification is a classic least squares regression problem once the dictionary is set and data is collected \cite{edmd, bishop2006pattern}.
While the pseudo inverse solution is straightforward, the matrix approximation can perform poorly especially in highly noisy data sets due to ill conditioned matrices.
On the other hand, probabilistic regression models can extract the desired information through the so called hidden variables and hyperparameters. 
In a probabilistic model with the observed data denoted as $\bx$, the hidden variables $\bz$ are unobserved variables in the model with generating parameters $\btheta$ and data generating probability density $p(\bx, \bz \mid \btheta)$ \cite{vb, bishop2006pattern}.

VB inference provides an approximate Bayesian treatment of probabilistic models with hidden variables by replacing the posterior $p(\bz \mid \bx, \btheta)$ with a known parameterized distribution family $q(\bz)$.
VB then connects naturally to the Expectation Maximization (EM) algorithm \cite{em} by maximizing the evidence lower bound
\begin{align}
    \LL(q, \btheta) & = \int q(\bz)\log\frac{p(\bx,\bz \mid \btheta)}{q(\bz)}\mathrm{d}\bz
\end{align}
in an alternating manner in E and M steps \cite{vb,bishop2006pattern}.
In particular, if the model lacks the generating parameters $\btheta$ and only contains hidden variables $\bz$, then the VB inference executes only the E step \cite{vb}.

In the presence of several hidden variables, the mean field approximation for the adopted distributions of the hidden variables can be implemented \cite{jordan1999introduction}:
\begin{align}
    q(\bz) = \prod_{i=1}^{N} q_i(z_i),
    \label{eq:mean_field_approximation}
\end{align}
where $N$ is the number of hidden variables.
Thus, the posterior distribution of each hidden variable is updated as \cite{vb}
\begin{align}
    \log q^*_i(z_i) =  \E_{\bz_{j\neq i}}[\log p(\bx, \bz)] + \text{const}.
    \label{eq:posterior_star}
\end{align}

\subsection{Condensation of a Graph}
Given a directed graph $G = (V, E)$ with the vertex set $V$ and the edge set $E$, a strongly connected component (SCC) is defined as a subset $S \subseteq V$ where every $u, v \in S$ are mutually reachable from another; i.e, directed paths exist between the two vertices in both directions.
SCCs constitute maximal subgraphs that are internally reachable.
Given a directed graph $G$, an associated condensed graph $G_C$ can be constructed by letting each SCC be a node.
The condensed graph $G_C$ of $G$ essentially captures the coarse interactions between these maximal subgraphs. 
Let $\{S_i\}_{i=1}^{N_S}$ be the SCCs of $G$.
We construct $G_C = (V_C, E_C)$ as $V_C = \cup_{i=1}^{N_S} \{S_i\}$ and $(S_i, S_j) \in E_C$ if and only if there exists an edge in $G$ from some vertex in $S_i$ to some vertex in $S_j$.
$G_C$ is consequently a coarser graph with the hierarchical structure preserved \cite{nuutila1994finding}. 

In \cite{graphdecomp}, the connection between graph representations of dynamical systems and SCCs has been laid out.
The so called sparsity graph $G$ of a dynamical system is created by associating the nodes with the states of a system, and constructing directed edges from node $i$ to $j$ if the dynamics of the state represented by $j$ depends on the state represented by $i$.
For each node $n$ without any outgoing edges in the condensation graph $G_C$ of $G$, the set of all nodes from which $n$ is reachable represents a dynamical subsystem \cite[Proposition 1]{graphdecomp}.

\section{Problem Statement}\label{sec:PS}

We consider a discrete time dynamical system described by $\bx[k+1] = \bS(\bx[k], \bu[k])$, where $\bx[k] \in \R^n$ is the state and $\bu[k] \in \R^l$ is the control input at time $k$.
We collect the consecutive measurements $\{ \bx[i] \}_{i=1}^{m+1}$ and $\{ \bu[i] \}_{i=1}^{m}$ from the system and we assume a dictionary of observables $\bphi = \begin{bmatrix} \phi_1 \ \ldots \ \phi_L \end{bmatrix}^T$ exists.
The goal is to find a subset $\bar{\bphi} \subseteq \cup \{\phi_i\}_{i=1}^L$ such that $\bar{\bphi}$ is the structurally minimal dictionary which includes all the observables affecting the output variables of the Koopman model.

We propose a probabilistic hierarchical dependency structure motivated by sparse dictionary selection under uncertainty.
To this end, we distinguish the effect of observable on the regression of another from its weight in the regression. 
The weight of an observable can carry physical meanings, have units and different scales of magnitude.
This ambiguity in interpreting the magnitude of a weight in the regression can render difficult for arbitrary dictionaries.
Therefore, the structural relevance of each observable should be represented separately from the numerical value of the corresponding regression coefficient.
Our goal requires inference of both the entries of $\hat{\bK}_F$ in \eqref{eq:koopman_model} and another matrix of the same size indicating whether the corresponding element should be included in the regression.
This indicator needs to be easily interpretable and dimensionless.
Probability values between $0$ and $1$ suit our objectives in this case.

We therefore adopt a probabilistic spike and slab type model \cite{spikeslab} and use VB inference to estimate the posterior distribution of model parameters.
In particular, we separate the effective weight of each regressor in two parts: One that describes the probability of inclusion, and one that describes the coefficient value conditional on inclusion. 
This separation is represented through the elementwise product $\gamma \odot \beta$, where $\gamma$ encodes whether the corresponding regressor is active and $\beta$ contains the associated coefficient values. 
The effective coefficient in the regression problem then becomes $\gamma \odot \beta$.
Conditional on the hyperparameters, we model $\gamma$ random variables as Bernoulli distributed and $\beta$ as Gaussian distributed.
This Bernoulli Gaussian construction induces a spike and slab type prior on the effective coefficient where inactive regressors have an exact zero contribution, whereas active regressors are assigned Gaussian distributed coefficients. 
In the end, the posterior inclusion probabilities $\E[\gamma]$ provide a dimensionless measure of dictionary relevance.
We also introduce coefficient precision hyperparameters to regulate shrinkage and noise precision to account for measurement uncertainty.
Lastly, we include the expectation of the inclusion flags as a random variable in the hierarchy so that dictionary pruning can be regulated.

Assuming the dynamical system has control inputs, we define the design matrix as $ {\bPhi = [ \bPhi' \mid \bU ] \in \R^{m \times (L+l)} }$ where $[\bPhi']_{ij} = \phi_j(\bx[i])$ for all $1 \leq i \leq m, 1 \leq j \leq L$ and $[\bU]_{ij} = u_j[i]$ for all $1 \leq j \leq l$.
Then, we treat the Koopman model regression of each dictionary element as an independent problem by setting the target regression variable $\bt_j = [\phi_j(\bx[2]) \dots \phi_j(\bx[m+1])]^T$ for each $1 \leq j \leq L$ so that
\begin{align}
    \bt_j = \bPhi (\bgamma_j \odot \bbeta_j)+ \bv_j \quad \bv_j \sim \N(\boldsymbol{0}, \rho_j^{-1}\eye_{m \times m}),
    \label{eq:wj_problem}
\end{align}
where $\bgamma_j$ is the vector of inclusion flags and $\bbeta_j$ is the vector of weights corresponding to each column of $\bPhi$, respectively. 
We then add a new level of hierarchy to our probabilistic model by assigning conditional priors to the coefficients $\bbeta_j$ and $\bgamma_j$.
In particular, we model $\bbeta_j$ as a vector of independent Gaussian random variables with zero mean and unknown precision as
\begin{align}
    \bbeta_j  & \sim \prod_{i=1}^{L+l} \N(\beta_{j,i}; \ 0, \ \alpha_{j,i}^{-1}).
    \label{eq:beta_cp} 
\end{align}
Then, we model the inclusion vector $\bgamma_j$ as independent Bernoulli distributions to represent its binary nature:
\begin{align}
    \bgamma_j & \sim \prod_{i=1}^{L+l} \text{Bernoulli}(\gamma_{j,i}; \pi_{j,i}).
    \label{eq;gamma_cp}
\end{align}
Finally, we complete the hierarchy by assigning priors to the hyperparameters in our model; namely, $\rho_j$, $\alpha_{j,i}$ and $\pi_{j,i}$ for each $j$ and $i$.
Gamma distribution, given by the shape and rate parameters, is the conjugate prior for the precision hyperparameter of Gaussian distribution \cite{bishop2006pattern}, so we assign $\rho_j$ and $\alpha_{j,i}$ Gamma distributions:
\begin{align}
    \rho_j    & \sim \text{Gamma}(\rho_j; \ a_j, \ b_j), \label{eq:rho_p}\\
    \balpha_j & \sim \prod_{i=1}^{L+l} \text{Gamma}(\alpha_{j,i}; \ c_{j,i}, \ d_{j,i}). \label{eq:alpha_p}
\end{align}
Similarly, the conjugate prior of Bernoulli distribution is the Beta distribution \cite{bishop2006pattern}. 
Consequently, we assign the prior distribution of $\pi_{j,i}$ as:
\begin{align}
    \bpi_j    & \sim \prod_{i=1}^{L+l} \text{Beta}(\pi_{j,i}; \ e_{j,i}, \ f_{j,i}). \label{eq:pi_p}
\end{align}
We estimate the posterior of all random variables by VB inference separately for each $ 1\leq j \leq L$.
The complete hierarchy for a given target variable is depicted in Figure \ref{fig:hierarchy}.
\begin{figure}
    \centering
    \tikzset{every picture/.style={line width=0.75pt}} %

\begin{tikzpicture}[x=0.75pt,y=0.75pt,yscale=-1,xscale=1]

\draw  [color={rgb, 255:red, 0; green, 0; blue, 0 }  ,draw opacity=1 ] (172.5,7.5) -- (229.5,7.5) -- (229.5,64.5) -- (172.5,64.5) -- cycle ;
\draw   (119.2,89.4) -- (169.2,89.4) -- (169.2,139.4) -- (119.2,139.4) -- cycle ;
\draw   (234.4,89.4) -- (284.4,89.4) -- (284.4,139.4) -- (234.4,139.4) -- cycle ;
\draw   (185.9,36) .. controls (185.9,27.66) and (192.66,20.9) .. (201,20.9) .. controls (209.34,20.9) and (216.1,27.66) .. (216.1,36) .. controls (216.1,44.34) and (209.34,51.1) .. (201,51.1) .. controls (192.66,51.1) and (185.9,44.34) .. (185.9,36)(181.3,36) .. controls (181.3,25.12) and (190.12,16.3) .. (201,16.3) .. controls (211.88,16.3) and (220.7,25.12) .. (220.7,36) .. controls (220.7,46.88) and (211.88,55.7) .. (201,55.7) .. controls (190.12,55.7) and (181.3,46.88) .. (181.3,36) ;
\draw   (128.95,114.4) .. controls (128.95,105.98) and (135.78,99.15) .. (144.2,99.15) .. controls (152.62,99.15) and (159.45,105.98) .. (159.45,114.4) .. controls (159.45,122.82) and (152.62,129.65) .. (144.2,129.65) .. controls (135.78,129.65) and (128.95,122.82) .. (128.95,114.4) -- cycle ;
\draw   (244.15,114.4) .. controls (244.15,105.98) and (250.98,99.15) .. (259.4,99.15) .. controls (267.82,99.15) and (274.65,105.98) .. (274.65,114.4) .. controls (274.65,122.82) and (267.82,129.65) .. (259.4,129.65) .. controls (250.98,129.65) and (244.15,122.82) .. (244.15,114.4) -- cycle ;
\draw   (268.6,17.2) -- (300.8,17.2) -- (300.8,49.4) -- (268.6,49.4) -- cycle ;
\draw   (125.8,154.8) -- (158,154.8) -- (158,187) -- (125.8,187) -- cycle ;
\draw   (243.8,154.2) -- (276,154.2) -- (276,186.4) -- (243.8,186.4) -- cycle ;
\draw    (145.8,89.2) -- (170.3,66.54) ;
\draw [shift={(172.5,64.5)}, rotate = 137.23] [fill={rgb, 255:red, 0; green, 0; blue, 0 }  ][line width=0.08]  [draw opacity=0] (8.93,-4.29) -- (0,0) -- (8.93,4.29) -- cycle    ;
\draw    (260.55,89.2) -- (231.85,66.37) ;
\draw [shift={(229.5,64.5)}, rotate = 38.5] [fill={rgb, 255:red, 0; green, 0; blue, 0 }  ][line width=0.08]  [draw opacity=0] (8.93,-4.29) -- (0,0) -- (8.93,4.29) -- cycle    ;
\draw    (141.8,154.95) -- (141.96,142.87) ;
\draw [shift={(142,139.88)}, rotate = 90.76] [fill={rgb, 255:red, 0; green, 0; blue, 0 }  ][line width=0.08]  [draw opacity=0] (8.93,-4.29) -- (0,0) -- (8.93,4.29) -- cycle    ;
\draw    (259.8,154.7) -- (259.76,142.87) ;
\draw [shift={(259.75,139.88)}, rotate = 89.81] [fill={rgb, 255:red, 0; green, 0; blue, 0 }  ][line width=0.08]  [draw opacity=0] (8.93,-4.29) -- (0,0) -- (8.93,4.29) -- cycle    ;
\draw    (269,34.38) -- (232.5,34.65) ;
\draw [shift={(229.5,34.67)}, rotate = 359.56] [fill={rgb, 255:red, 0; green, 0; blue, 0 }  ][line width=0.08]  [draw opacity=0] (8.93,-4.29) -- (0,0) -- (8.93,4.29) -- cycle    ;

\draw (197.2,30.6) node [anchor=north west][inner sep=0.75pt]    {$\bt$};
\draw (138.4,106.6) node [anchor=north west][inner sep=0.75pt]    {$\bbeta $};
\draw (254.4,108.6) node [anchor=north west][inner sep=0.75pt]    {$\bgamma $};
\draw (278.8,27.4) node [anchor=north west][inner sep=0.75pt]    {$\rho $};
\draw (136,166) node [anchor=north west][inner sep=0.75pt]    {$\balpha $};
\draw (252.8,166) node [anchor=north west][inner sep=0.75pt]    {$\bpi $};

\end{tikzpicture}
    \caption{The hierarchy of the random variables in the proposed model.}
    \label{fig:hierarchy}
\end{figure}
Once the estimation of random variables is complete, we use the estimates of the random variables to construct $\hat{\bK}_F = \begin{bmatrix} \hat{\bgamma}_1 \odot \hat{\bbeta}_1 \mid \dots \mid \hat{\bgamma}_L \odot \hat{\bbeta}_L \end{bmatrix}$ and obtain the probabilistic Koopman model:
\begin{align}
    \bphi(\bx[k+1])^T = \begin{bmatrix} \bphi(\bx[k])^T \bu[k]^T \end{bmatrix} \hat{\bK}_F + \bm{v}^T 
\end{align}
where $\bm{v} \sim \N(\boldsymbol{0} , \diag(\hat{\rho}_1^{-1}, \dots, \hat{\rho}_L^{-1}))$ is a plug in predictor of the regression covariance.
As the last step of our task, we utilize the so called inclusion matrix $\bGamma = \begin{bmatrix}
    \hat{\bgamma}_1 \mid \dots \mid \hat{\bgamma}_L
\end{bmatrix}$ to find $\bar{\bphi}$.

\section{Method}\label{sec:method}
\subsection{VB inference of posterior distributions}\label{sec:VBinf}
The equations \eqref{eq:wj_problem}--\eqref{eq:pi_p} set up the prior model. 

We then seek the estimates of the posterior distributions of each random variable in each of the $L$-many independent problems.
Since the problems are identical for each of the $L$-many targets variables $\{\bt_j\}_{j=1}^L$, we drop the subscript $j$ in all the random variables for brevity. 
The posterior estimation approach allows our model to take shape according to data and the prior beliefs.
In this regard, parameters in the prior model allow representation of initial beliefs about the random variables. 
Typically, Koopman model dictionaries are redundant.
Therefore, sparsity promoting parameters represent our belief about the redundancy.
In particular, we set the prior mean of $\beta_i$ as $0$.
Similarly, initial $\pi_i$ is set as $0.5$ to enable flexibility and we set $e_i$ and $f_i$ such that the mass of Beta distribution is concentrated near 0 when sparsity is expected.
Alternatively, a uniform Beta prior is also possible. 

The exact posterior distribution of all parameters is costly to find.
We therefore treat the random variables in our model as hidden variables and resort to the mean field approximation \eqref{eq:mean_field_approximation}.
Accordingly,
\begin{align}
    p(\bbeta, \bgamma, \balpha, \bpi, \rho \mid \bt ) \approx q(\rho) \prod_{i=1}^{L+l} q(\beta_i) q(\gamma_i) q(\alpha_i) q(\pi_i),
\end{align}
where each $q(\cdot)$ inherits the corresponding distribution type of the same random variable in our model prior.
This means for each $1 \leq i \leq (L+l)$, $q(\beta_i)$, $q(\gamma_i)$, $q(\alpha_i)$ and $q(\pi_i)$ are Gaussian, Bernoulli, Gamma and Beta distributions, respectively.
Likewise, $q(\rho)$ is a Gamma distribution.
Consequently, \eqref{eq:posterior_star} applies for posterior estimation.
Our model prior allows the conditional decomposition of the joint distribution as
\begin{align}
    \begin{split}
        &p(\bt, \bbeta, \bgamma, \balpha, \bpi, \rho) = \\ &p(\bt \mid \bbeta, \bgamma, \rho) \ p(\bbeta \mid \balpha) \ p(\bgamma \mid \bpi) \ p(\balpha) \ p(\bpi) \ p(\rho).
    \end{split}
    \label{eq:decomp_prior}
\end{align}
The VB inference procedure starts from an initial set $\left\{ a, b, \left\{ c_i, d_i, e_i, f_i, \alpha_i, \mu_i, \pi_i\right\}_{i=0}^{L+l} \right\}$ for the model parameters.
Then, we perform updates iteratively in the Jacobi fashion to allow for parallelization \cite{saad2003iterative}.
An overview of the algorithm is given in Algorithm \ref{alg:SpikeAndSlab} where all right hand side expectations are evaluated using the variational factors from the previous iteration. The new factors are committed after completing the updates.

We denote by the barred variables the posterior parameter estimates after the update takes place and by hat variables the expectations of the random variables after the previous iteration.
The symbol $\peq$ is used to indicate that the equation holds up to an additive constant. 

We provide the detailed derivations of the update rules of the parameters in Appendix \ref{sec:apA} for brevity.
Here, we present the final update formulas for each parameter.
In our derivations, the regressors $\phi_i$ correspond to the observables when $1 \leq i \leq L$ and to the control inputs when $L < i \leq L+l$.
\subsubsection{Update of $q(\rho)$}
Given prior parameters $a$ and $b$ for $p(\rho)$, the posterior update of $q(\rho)$ is Gamma distributed with the shape and rate parameters $\bar{a}$ and $\bar{b}$, respectively:
\begin{align}
\begin{split}
   \bar{a} &= \frac{m}{2} + a,\\
   \Bar{b} = \frac{1}{2} \bigg[ ||\bt - &\Phi(\hat{\bgamma} \odot \bmu)||^2\\ + \sum_{i=1}^{L+l} ||\bphi_{i}||^2 \big( \hat{\gamma}_i (&\mu_i^2 + \sigma_i^2) - \hat{\gamma}_i^2\mu_i^2 \big) \bigg] +  b.    \label{eq:a_b_update}
\end{split}
\end{align}
\subsubsection{Update of $q(\alpha_i)$}
Given the prior parameters $c_i$ and $d_i$ for $p(\alpha_i)$, the posterior $q(\alpha_i)$ is Gamma distributed with shape and rate parameters $\bar{c}_i$ and $\bar{d}_i$, respectively:
\begin{equation}
         \bar{c}_i = c_i + \frac{1}{2}, \ \bar{d}_i = d_i + \frac{\mu_i^2+ \sigma_i^2}{2}, \label{eq:c_d_update}    
\end{equation}
where $\mu_i$ is the posterior mean and $\alpha_i$ is the posterior precision of $q(\beta_i)$ from the previous iteration.
\subsubsection{Update of $q(\pi_i)$}
Given the prior parameters $e_i$ and $f_i$ of $p(\pi_i)$, the posterior $q(\pi_i)$ is Beta distributed with parameters $\bar{e}_i$ and $\bar{f}_i$, respectively:
\begin{equation}
          \bar{e}_i = \hat{\gamma}_i + e_i, \ \ \bar{f}_i = 1 - \hat{\gamma}_i + f_i. \label{eq:e_f_update}   
\end{equation}
\subsubsection{Update of $q(\beta_i)$}
In the Gaussian prior $p(\beta_i)$, we set the mean to $0$ and the precision to $\alpha_i$.
The posterior $q(\beta_i)$ is Gaussian distributed with mean $\mu_i$ and precision $\bar{\alpha}_i$.
We define the residual of the $k$-th target measurement without the $i$-th regressor:
\begin{equation}
    r_{ik} = \bigg(t_k - \sum_{ \substack{j=1 \\ j\neq i}}^{L+l}\gamma_j\beta_j\phi_j(\bx[k]) \bigg).
    \label{eq:r_ik}
\end{equation}
The posterior $q(\beta_i)$ is then Gaussian distributed with mean $\bar{\mu}_i$ and precision $\bar{\alpha}_i$:
\begin{subequations}
    \begin{align}
         \bar{\alpha}_i = \sum_{k=1}^{m} \hat{\rho}\hat{\gamma}_i \phi_i^2(\bx[k]) + \hat{\alpha}_i, \\
         \bar{\mu}_i = \frac{\sum_{k=1}^{m} \hat{\rho}\hat{\gamma}_i\phi_i(\bx[k]) \hat{r}_{ik}}{\sum_{k=1}^{m} \hat{\rho}\hat{\gamma}_i \phi_i^2(\bx[k]) + \hat{\alpha}_i}.        
    \end{align}
    \label{eq:alpha_mu_update}
\end{subequations}
We also define and use the posterior variance of $\beta_i$ as $\sigma_i^2 := \bar{\alpha}_i^{-1}$.

In our implementations, we have seen the benefit of introducing damping to the updates for $\mu_i$ and $\alpha_i$. 
This technique is used in the literature where the loss surface is rugged \cite{Vu_2024}.
With the introduction of damping coefficient $p_d \in (0,1)$, updates become:
\begin{subequations}
    \begin{align}
    \bar{\alpha}_{s,\mathrm{damped}}
    &= p_d \bar{\alpha}_s
    + (1-p_d)\bar{\alpha}_{s-1,\mathrm{damped}}, \\
    \bar{\mu}_{s,\mathrm{damped}}
    &= p_d \bar{\mu}_s
    + (1-p_d)\bar{\mu}_{s-1,\mathrm{damped}}.
    \end{align}
    \label{eq:damping}
\end{subequations}
where \(s\) is the iteration index.
\subsubsection{Update of $q(\gamma_i)$}
We first stack $r_{ik}$ in a vector:
\begin{equation}
    \bm{r}_i = \begin{bmatrix}
                   r_{i1} & \dots & r_{im}
               \end{bmatrix}^T \ \text{for} \ i=1, \dots, L+l.
               \label{eq:br}
\end{equation}
Similarly, we define for all $1 \leq i \leq L+l$
\begin{align}
        \bphi_i = \begin{bmatrix}
                  \phi_i(\bx[1]) & \dots & \phi_i(\bx[m])
              \end{bmatrix}^T.
\end{align}
Then we define the exponent
\begin{equation}
 \eta_i =\hat{\rho}\mu_i\bm{\phi}_i^T\hat{\bm{r}}_i - \frac{1}{2}\hat{\rho} (\mu_i^2 + \sigma_i^2) ||\bm{\phi}_i||^2 + \psi(\bar{e}_i) - \psi(\bar{f}_i),
 \label{eq:eta_defn}
\end{equation}
where $\psi(\cdot)$ is the Digamma function \cite{spouge1994computation}.
Finally, the posterior $q(\gamma_i)$ is Bernoulli distributed with expectation $\bar{\pi}_i$:
 \begin{equation} \label{eq:pi_update}
     \bar{\pi}_i = \frac{e^{\eta_i}}{1 + e^{\eta_i}} = \frac{1}{1+ e^{-\eta_i}} = \sigma(\eta_i),
 \end{equation}
 where $\sigma(\cdot)$ is the sigmoid function \cite{bishop2006pattern}.
Here, we have experimentally observed the improved numerical stability when we clip $\bar{\pi}_i$ between $[\delta, 1-\delta]$ where $\delta$ is a small number, such as $10^{-8}$.

\begin{algorithm}[t]
    \caption{VB Inference for the Proposed Model for a Given Target Type $k$}
    \label{alg:SpikeAndSlab}
    \begin{algorithmic}[1]
        \State \textbf{Input:} $\Phi$, $\bt$, $a$, $b$, $\{c_i\}_{i=1}^{L+l}$, $\{d_i\}_{i=1}^{L+l}$, $\{e_i\}_{i=1}^{L+l}$, $\{f_i\}_{i=1}^{L+l}$, $p_d$
        \State Initialize $q(\beta_i), q(\gamma_i), q(\alpha_i), q(\pi_i), q(\rho)$ for all $i$.
        \Repeat
        \State Update $q(\rho)$ using \eqref{eq:a_b_update}
        \For{$i = 1$ to $L+l$}
        \State Update $q(\alpha_i)$ using \eqref{eq:c_d_update}
        \State Update $q(\pi_i)$ using \eqref{eq:e_f_update}
        \State Compute the residual using \eqref{eq:r_ik}
        \State Update $q(\beta_i)$ using \eqref{eq:alpha_mu_update}
        \State Damping update for \(q(\beta_i)\) using \eqref{eq:damping}
        \State Update $q(\gamma_i)$ using \eqref{eq:eta_defn}, \eqref{eq:pi_update} \
        \EndFor
        \For{$i = 1$ to $L+l$}
        \State Update the expectations:
        \State \(\hat{\rho} = \frac{\bar{a}}{\bar{b}}\)
        \State \(\hat{\alpha}_i = \frac{\bar{c}_i}{\bar{d}_i}\)
        \State \(\hat{\pi}_i = \frac{\bar{e}_i}{\bar{e}_i + \bar{f}_i}\)
        \State \( \sigma^2_i = \bar{\alpha}_{i-\mathrm{damped}}^{-1} \)
        \State \( q(\beta_i) = \N(\bar{\mu}_{i-\mathrm{damped}}, \sigma^2_i)\)
        \State \(\hat{\gamma_i} = \bar{\pi}_i \)
        \EndFor
        \Until{Maximum number of iterations is reached or converged}
        \State \textbf{Output:} Posteriors $q(\bbeta), q(\bgamma), q(\balpha), q(\bpi), q(\rho)$
    \end{algorithmic}
\end{algorithm}

\subsubsection{Time complexity of VB inference}
We define \(p = L + l\).
The complexity of a single iteration of VB inference per target is $\mathcal{O}(mp)$ when the residual is cached.
For $L$ targets, the serial complexity of one iteration of VB becomes $O(mpL)$.
Usually, $l<<L$ for large Koopman dictionaries, corresponding to a complexity level of $\mathcal{O}(mL^2)$ which is linear in measurements and quadratic in Koopman dictionary size.

\subsection{Dictionary Reduction Using the Inclusion Matrix}
The posterior estimates of the inclusion random variables in \eqref{eq:wj_problem} represent the probabilities of
participation for each regressor in each observable regression. 
Once the VB iterations are completed for all target observables, we obtain the inclusion
matrix
\begin{align}
    \bGamma = [\hat{\bgamma}_1 \; | \; \cdots \; | \; \hat{\bgamma}_L]
    \in [0,1]^{(L+l)\times L}.    
\end{align}
When a regressor is present in the regression for a target variable, the state transition of the target depends on the regressor.
The first $L$ rows of $\bGamma$ correspond to observable to observable
dependencies, while the last $l$ rows correspond to control input to observable dependencies. 
In this subsection, we draw similarities with the dynamical system representation in \cite{graphdecomp} and the implications of our hierarchical model.
In particular, we use $\bGamma$ to adapt a graph representation for the dynamical system and reduce the state dimension of the Koopman model.
Since the control input is exogenous and is not part of the observable dictionary to be pruned, the graph construction proposed here is based on the
observable block of $\bGamma$, denoted by
\begin{align}
    \bGamma_x := \bGamma_{1:L,1:L}.
\end{align}
For a prescribed probability threshold $\epsilon \in (0,1)$, we define the
binary matrix
\begin{align}
    [\bGamma_x^\epsilon]_{ij} = \begin{cases}
        0, \ \text{if} \ [\bGamma_x]_{ij} \leq \epsilon \\
        1, \ \text{if} \ [\bGamma_x]_{ij} > \epsilon
    \end{cases}. 
    \label{eq:threshold_gamma}
\end{align}
This thresholding step converts posterior inclusion probabilities into a sparsity pattern where a regressor is discarded if it has less than $\epsilon$ probability to be included in the regression of a target. 
Equivalently, it defines the thresholded finite dimensional Koopman model
\begin{align}
    \bphi[k+1]^{T}=\bphi[k]^{T}\mathbf{A}_\epsilon^T+\bu[k]^{T}\mathbf{B}^T,
\label{eq:thresholded_koopman_model}
\end{align}
where $\mathbf{A}_\epsilon^T \in \R^{L\times L}$ has the sparsity pattern of
$ \bGamma_x^\epsilon $.
The entry $[\mathbf{A}_\epsilon^T]_{ij}$ contributes to the one step evolution of $\phi_j$ from $ \phi_i $. 
Therefore, \([\bGamma_x^\epsilon]_{ij}=1\) means that observable
\(\phi_i\) is retained as a regressor for the dynamics of observable \(\phi_j\).

We associate \(\bGamma_x^\epsilon\) with a directed graph
\[
    G^\epsilon=(V,E), \qquad V=\{1,\ldots,L\},
\]
where the node $i$ represents the observable $\phi_i$, and edge
$(i,j)\in E$ if and only if $[\bGamma_x^\epsilon]_{ij}=1$. 
Thus, an edge $(i,j)$ represents a direct one step dependence of the Koopman state
$\phi_j$ on $\phi_i$. 
A directed path from $i$ to $j$ represents an indirect multi step influence. 
This is the finite dimensional Koopman analogue
of the sparsity graph used for subsystem decomposition in \cite{graphdecomp},
where the graph encodes causal dependencies between dynamical states.
In \cite{graphdecomp}, the sparsity graph of a dynamical system represents causal dependence between state dynamics; the past of a node is shown to characterize the smallest subsystem containing that node, and SCC condensation is used to group mutually dependent states into indivisible subsystem blocks. 
Here, the same graph theoretic principle is specialized to the finite dimensional Koopman representation by interpreting $\bGamma_x^\epsilon$ as the sparsity graph of the lifted dynamics.

In the following, we define the ancestors of a node, which corresponds to the past of a node in \cite{graphdecomp}.

\begin{definition}
We call a node $i$ an ancestor of node $j$ if there exists a directed path from $i$ to $j$.
Trivially, every node is an ancestor of itself.
For a given node, $\A(j)$ denotes the set of ancestors of the node $j$.
For a set of nodes $S \subseteq V$, $\A(S)$ is given as
\begin{align}
    \A(S) = \bigcup_{j\in S}\mathcal{A}(j).
\end{align}    
For ancestors in a condensation graph, we use the notation $A_C(\cdot)$.
\end{definition}

Let $O\subseteq V$ denote the index set of the output observables. 
In this work, $O$ typically contains the observables corresponding to the measured
physical states. 
We define $R = \A(O)$ as the ancestors of the Koopman model output nodes. 
The ancestor set $R$ can be computed directly on
$G^\epsilon$. 
However, if two or more observables are mutually reachable,
they belong to the same SCC. 
In dynamical terms, such observables are mutually dependent through the thresholded Koopman dynamics and should be treated as a single block.
Therefore, following subsystem decomposition arguments in \cite{graphdecomp}, we first condense the graph by contracting each SCC into a single node. The resulting condensation graph is acyclic and preserves the reachability relation between blocks.
In the following, we show that $R$ can be equivalently computed on the condensation graph $G_C^\epsilon$ of $G^\epsilon$.

\begin{definition}[]
    Let $G^\epsilon = (V,E)$ be the Koopman dependency graph and let $\mathcal{S} = \{S_1, \dots, S_M\}$ denote its SCCs.
    Let $G_C^\epsilon = (\mathcal{S},E_C)$ denote its condensation graph.
    For a set of components $\mathcal{Q} \subseteq \mathcal{S}$, we define the decondensation map 
    \begin{align*}
        \operatorname{dec}({\mathcal{Q}}) := \{v\in V \;|\; v\in S \text{ for some } S\in\mathcal{Q}\}.
    \end{align*}
\end{definition}

In the next proposition, we look into the equivalence of ancestry search in the original and the condensed graphs.

\begin{proposition}[Equivalence of ancestry search before and after SCC condensation]\label{prop:scc_equivalence}
    For the output index set $O$ and the set of SCCs $ \mathcal{S}_O$ containing the output vertices, the following holds:
    \begin{align*}
        \A(O) = \operatorname{dec}(\A_C(\mathcal{S}_O)),        
    \end{align*}
    where $\A(O)$ denotes the ancestor set of $O$ in $G^\epsilon$,
    and $\A_C(\mathcal{S}_O)$ denotes the ancestor set of
    $\mathcal{S}_O$ in the condensation graph $G_C^\epsilon$.
\end{proposition}
\begin{proof}
    We show both inclusions.
    
    $(\subseteq)$ First, let $v \in \A(O)$.
    Then, there exist an output node $o \in O$ and a directed path from $v$ to $o$ in $G^\epsilon$. 
    Let $S_v$ and $S_o$ be the SCCs containing $v$ and $o$, respectively. 
    We replace each node $i$ in the path from $v$ to $o$ with the corresponding SCC $S_i$ such that $i \in S_i$. 
    By removing the repeated SCCs if necessary, we get a directed path from $S_v$ to $S_o$. 
    As a result, $S_v \in \A_C(S_o)$ and $v \in \operatorname{dec}(\A_C(S_o))$.

    $(\supseteq)$ Next, let $v \in \operatorname{dec}(\A_C(\mathcal{S}_O))$.
    Then, $v \in S_v$ for some SCC $S_v \in \A_C(\mathcal{S}_O)$. 
    Consequently, there exists $S_o \in \mathcal{S}_O$ such that there is a directed path from $S_v$ to $S_o$. 
    By construction, each edge between two SCCs in  $G^\epsilon_C$ corresponds to at least one edge in $G^\epsilon$ between nodes belonging to these SCCs. 
    Since nodes in SCCs are mutually reachable, the directed path in SCC level can be lifted to a directed path in  $G^\epsilon$ from $v$ to some node $o \in S_o\cap O$.
    As a result, $v \in \A(O)$.
\end{proof}

We show next that $R$ is closed under the dependencies induced by the Koopman model.

\begin{lemma}[Koopman dependency closure]\label{lemma:dependency_closure}
The set of ancestors of the output nodes $R = \A(O)$ is closed with respect to the incoming dynamical dependencies.
In other words, if $j\in R$ and $(i,j)\in E$, then $i\in R$.
\end{lemma}

\begin{proof}
Since $j\in R$, there exists an output vertex $o\in O$ such that $j$ is an
ancestor of $o$. 
Equivalently, there exists a directed path from $j$ to
$o$, or $j=o$. 
If $(i,j)\in E $, then concatenating the edge $(i,j)$ with
the path from $j$ to $o$ gives a directed path from $i$ to $o$.
Hence, $i$ is also an ancestor of an output vertex, and therefore $i\in R$ by definition.
\end{proof}

Lemma \ref{lemma:dependency_closure} states that no observable that is not in $R$ can directly enter the
one step dynamics of any observable in $R$. 
This property is the key
condition required for output preserving reduction.

The next proposition shows that the trajectory of the output observables are only dependent on the observables in $R$.
We call $R$ to be the retained set of observables and show it is an output preserving reduction of the initial dictionary.

\begin{proposition}[Output preservation of the reduced Koopman model]\label{prop:reduction}
Consider the thresholded Koopman model \eqref{eq:thresholded_koopman_model}, the retained observables $R=\A(O)$. 
Let $D=V\setminus R$ be the discarded set. 
Then, for any initial lifted state $\bphi[0]$ and any input sequence $\{\bu[k]\}_{k\geq 0}$, the trajectory of the retained coordinates $\bphi_R[k]$ generated by the full thresholded model is identical to the trajectory generated by the reduced model
\begin{align}
    \bphi_R[k+1]^T = \bphi_R[k]^T \mathbf{A}_{\epsilon RR}^T
    + \bu[k]^T \mathbf{B}_R^T .   
\end{align}
Consequently, the output coordinates indexed by $ O\subseteq R $ are preserved
exactly by the reduction.
\end{proposition}
\begin{proof}
Permute the observable vector as $\bphi=[\bphi_R^T \; \bphi_D^T]^T$. 
With the same partitioning, the thresholded Koopman matrix can be written as
\begin{align}
\mathbf{A}_\epsilon ^T =
\begin{bmatrix}
\mathbf{A}_{\epsilon RR}^T & \mathbf{A}_{\epsilon RD}^T\\
\mathbf{A}_{\epsilon DR}^T & \mathbf{A}_{\epsilon DD}^T
\end{bmatrix},    
\label{eq:Ablocks}
\end{align}

The update of the retained coordinates is therefore
\[
    \bphi_R[k+1]^T = \bphi_R[k]^T \mathbf{A}_{\epsilon RR}^T + \bphi_D[k]^T \mathbf{A}_{\epsilon DR}^T
    + \bu[k]^T \mathbf{B}_R^T .
\]
By Lemma \ref{lemma:dependency_closure}, if $j\in R$ and $i\in D$, then the edge
$(i,j)$ cannot exist. 
Hence, $[\mathbf{A}_\epsilon]^T_{ij}=0$ for every
$i\in D$ and $j\in R$, which implies $\mathbf{A}_{\epsilon DR}^T=0$. Thus,
\[
    \bphi_R[k+1]^T = \bphi_R[k]^T \mathbf{A}_{\epsilon RR}^T
    + \bu[k]^T \mathbf{B}_R^T . 
\]
The retained dynamics are independent of $\bphi_D$. 
By induction on $k$, the retained trajectory generated by the full model and the reduced model is identical for the same initial $\bphi_R[0]$ and the same input sequence. 
Since $O\subseteq R$, all output observables are included in the retained state, and
their predictions are therefore preserved.
\end{proof}
\begin{remark}
Proposition \ref{prop:reduction} shows that the observables kept in $R$ constitute a dictionary that preserves the output.
While the proposition is an exact statement for the thresholded finite dimensional Koopman model, 
the relationship to the true nonlinear system remains data dependent, because $\bGamma_x^\epsilon$ is inferred from finite and noisy data. 
Therefore, the threshold $\epsilon$ determines a model selection tradeoff: a small value of $\epsilon$ retains weak dependencies and reduces the risk of removing relevant observables, while a large value of $\epsilon$ yields a more aggressive reduction. 
\end{remark}
We next show that $R = \A(O)$ is the smallest set which is closed under the Koopman model dependencies of $O$.

\begin{lemma}[Minimality of the ancestor reduction]\label{lemma:minimality}
Let \(S\subseteq V\) be any set of observable indices such that \(O\subseteq S\)
and \(S\) is closed under incoming dependencies; that is, if \(j\in S\) and
\((i,j)\in E\), then \(i\in S\). Then $\mathcal{A}(O)\subseteq S$.
\end{lemma}
\begin{proof}
Let $i_0 \in \A(O)$. Then there exists $o\in O$ and a directed path
$i_0\to i_1\to \cdots \to i_r=o$. Since $o\in O\subseteq S$, and $S$ is closed under incoming dependencies, $i_{r-1}\in S$. Repeating the same argument backward along the path inductively gives $i_{r-2} \ldots,i_0 \in S$.
Thus, every ancestor of every output node must be contained in any dependency closed set containing the outputs. 
\end{proof}
\begin{remark}
    Lemma \ref{lemma:minimality} ensures that if there is a subset $S\subseteq \A(O)$ such that $O \subseteq S$, and S is closed under incoming dependencies, then $S = \A(O)$.
\end{remark}
Lemma \ref{lemma:minimality} shows that the ancestor set is the smallest dependency closed set that contains the output observables.

Our findings demonstrate that the outputs of the Koopman model are preserved by dictionary reduction.
This property should be distinguished from controllability preservation.
Finite dimensional Koopman models are linear predictors in a lifted observable space, and controllability of the identified pair is not guaranteed by Koopman lifting \cite{miao2026learning}.
Likewise, the proposed VB inference procedure is not intended to preserve controllability. 
Therefore, the proposed reduction is not intended as a general controllability preserving model reduction for arbitrary lifted states.
Nevertheless, because the ancestry search yields a retained subsystem which is closed with respect to incoming dependencies, it preserves the controllable directions of the full lifted model after projection onto the retained coordinates, as stated next.

\begin{proposition}[Controllable subspace preservation under dependency closed reduction] \label{prop:controllability}
Consider the thresholded lifted Koopman model \eqref{eq:thresholded_koopman_model}
and let $R=\A(O)$ be the retained observables and $D=V\setminus R$ be the discarded observables. 
Since $R$ is dependency closed, we permute the coordinates as $\bphi=[\bphi_R^T \bphi_D^T]^T$ and write $\mathbf{A}^T_\epsilon$ as in \eqref{eq:Ablocks} with $[\mathbf{A}^T_{\epsilon DR}]_{ij}=0$ by Lemma \ref{lemma:dependency_closure}.
Further, we permute $\mathbf{B}^T$ if necessary to obtain $\mathbf{B}^T = \begin{bmatrix}
    \mathbf{B}^T_R & \mathbf{B}^T_D 
\end{bmatrix}$.
In the standard column vector convention, this gives the block structure
\begin{align}
    \mathbf{A}_\epsilon = \begin{bmatrix}
        \mathbf{A}_{\epsilon RR} & 0 \\ \mathbf{A}_{\epsilon RD} & \mathbf{A}_{\epsilon DD}
    \end{bmatrix}, \ 
    \mathbf{B} = \begin{bmatrix}
        \mathbf{B}_R \\ \mathbf{B}_D 
    \end{bmatrix}.
\end{align}
Let
\begin{align*}
    \mathcal{C}_{\mathrm{full}}
    =
    \operatorname{span}\{\mathbf{B},\mathbf{A}_\epsilon \mathbf{B},\ldots,(\mathbf{A}_\epsilon)^{L-1}\mathbf{B}\}    
\end{align*}
be the controllable subspace of the full lifted model, and let
\begin{align*}
    \mathcal{C}_{R}
    =
    \operatorname{span}\{\mathbf{B}_R,\mathbf{A}_{\epsilon RR} \mathbf{B}_R,\ldots,
    (\mathbf{A}_{\epsilon RR}^{|R|-1})\mathbf{B}_R\}    
\end{align*}
be the controllable subspace of the reduced model. 
Then
\begin{align*}
    \bm{\Pi}_R \mathcal{C}_{\mathrm{full}}=\mathcal{C}_{R},
\end{align*}
where $\bm{\Pi}_R$ denotes projection onto the retained coordinates.
\end{proposition}
\begin{proof}
Let $\bm{\Pi}_R$ be the projection that recovers the retained coordinates. 
The block triangular structure gives
\begin{align*}
 \bm{\Pi}_R \mathbf{A}_\epsilon \begin{bmatrix}
     \bphi_R \\ \bphi_D
 \end{bmatrix} &= \mathbf{A}_{\epsilon RR}\bm{\Pi}_R  \begin{bmatrix}
     \bphi_R \\ \bphi_D
 \end{bmatrix}, \\
  \bm{\Pi}_R \mathbf{B} &= \mathbf{B}_R.
\end{align*}
We now show by induction that for all $k \geq 0$, 
\[\bm{\Pi}_R \mathbf{A}_\epsilon^k \mathbf{B} = \mathbf{A}_{\epsilon RR}^k \mathbf{B}_R.\]
For $k = 0$, we have $\bm{\Pi}_R \mathbf{B} = \mathbf{B}_R$.
Assume the identity holds for some $k\geq0$.
Then,
\begin{align*}
    \bm{\Pi}_R \mathbf{A}_\epsilon^{k+1} \mathbf{B} &= \mathbf{A}_{\epsilon RR} \bm{\Pi}_R \mathbf{A}_\epsilon^k \mathbf{B} \\
    &= \mathbf{A}_{\epsilon RR} \mathbf{A}_{\epsilon RR}^k \mathbf{B}_R. \\
    &= \mathbf{A}_{\epsilon RR}^{k+1} \mathbf{B}_R.
\end{align*}
Hence the identity holds for all $k \geq 0$ by induction.

Therefore, projecting the columns of the full controllability matrix onto the retained coordinates gives exactly the columns of the reduced controllability matrix:
\begin{align*}
    \bm{\Pi}_R
    \begin{bmatrix}
    \mathbf{B} & \mathbf{A}_{\epsilon} \mathbf{B} & \cdots & \mathbf{A}_{\epsilon}^{L-1} \mathbf{B}
    \end{bmatrix}\\
    =
    \begin{bmatrix}
    \mathbf{B}_R & \mathbf{A}_{\epsilon RR} \mathbf{B}_R & \cdots &
    \mathbf{A}_{\epsilon RR}^{L-1} \mathbf{B}_R
    \end{bmatrix}.    
\end{align*}
Since the Cayley-Hamilton theorem implies that powers beyond \(|R|-1\) do not increase the reduced controllable subspace, the projected full controllable subspace equals $\mathcal{C}_R$. 
The remaining claims follow immediately.
\end{proof}

\begin{algorithm}[t]
\caption{Dictionary Reduction via Koopman Dependency Graph}
\label{alg:dictionary_reduction}
\begin{algorithmic}[1]
\State \textbf{Input:} Koopman model \ref{eq:koopman_model}, posterior inclusion matrix for observables $\bGamma_x=\bGamma_{1:L,1:L}$, output
index set $O$, threshold $\epsilon \in (0,1)$
\State Threshold \(\bGamma_x\) using \eqref{eq:threshold_gamma} to obtain \(\bGamma_x^\epsilon\).
\State Construct the directed graph $G^\epsilon=(V,E)$, where
$V=\{1,\ldots,L\}$ and $(i,j)\in E$ if
$[\bGamma_x^\epsilon]_{ij}=1$.
\State Compute the SCCs $\{S_i\}_{i=1}^{M}$ of $G^\epsilon$.
\State Construct the condensation graph $G_C^\epsilon = (\{S_i\}_{i=1}^{M}, E_C)$.
\State Identify the SCCs $\mathcal{S}_O$ that contain the output nodes $O$.
\State Compute the ancestors of the SCCs containing the output nodes $\A_C(\mathcal{S_O)}$ in the condensation graph.
\State Using the decondensation map , let $R= \operatorname{dec}(\A_C(\mathcal{S}_O))$.
\State Construct the reduced dictionary
\[
    \bar{\bphi}=\{\phi_i \;|\; i\in R\}.
\]

\State \textbf{Output:} Reduced dictionary $\bar{\bphi}$
\end{algorithmic}
\end{algorithm}

As a result of our theoretical findings, we prune the initial dictionary such that only the observables in $R$ are kept.
By Proposition \ref{prop:scc_equivalence}, we perform graph search on the condensed graph $G_C$ to obtain $\A_C(S_O)$.
Then, we use the decondensation map to find the retained observables.
Finally, our workflow includes using $\bar{\bphi} = \{\phi_i \;|\; i\in R\} $ to construct a reduced Koopman model. We then estimate the parameters of the reduced model from data once again.
We present the overall dictionary reduction procedure in Algorithm \ref{alg:dictionary_reduction}.

\begin{remark}
The theoretical findings we propose are exact for the thresholded finite dimensional models and the corresponding block restricted subsystems. 
In the numerical experiments, after the reduced dictionary is selected, the Koopman matrices are identified again from measured data using the selected dictionary. 
This refitting step is useful for numerical consistency and allows the reduced dictionary to be combined with different Koopman identification methods.
However, the relearned reduced model is no longer exactly the block restriction of the original thresholded model. 
Therefore, differences between the full and reduced model predictions are expected. The exact output preservation and controllable subspace identities should be interpreted as properties of the thresholded block restricted model rather than of every subsequently refit reduced model.
\end{remark}

Computational complexity of the matrix thresholding operation, as presented in Algorithm \ref{alg:dictionary_reduction}, is $\mathcal{O}(L^2)$ and ancestry search in $G_C^\epsilon$ has complexity $\mathcal{O}(M + |E_C|)$ \cite{tarjan1972depth}.
Since $|E_C| \leq M^2$ and $M \leq L$, Algorithm \ref{alg:dictionary_reduction} has an overall complexity of $O(L^2)$.
Lastly, let $r = |R|$. 
Before dictionary reduction, the Koopman model stores $\hat{\bK}_F \in \R^{(L+l)\times L}$, hence, the number of stored coefficients is then $N_{\mathrm{full}} = L(L+l)$.
After model dimension reduction, only $r \leq L$ observables are retained, the reduced model thereby stores $N_{\mathrm{red}} = r(r+l)$ coefficients. 
Therefore the number of stored coefficients is quadratic in the dictionary size.

\section{Results}\label{sec:results}
\subsection{Prediction performance}
We compare the results of various Koopman EDMD identification methods with full sized and reduced dictionaries across a representative spectrum of systems.
Even though we provide the results of the proposed method in the reduced models, we present our algorithm as a dictionary reduction framework to be used in conjunction with various identification methods rather than a standalone identification scheme. 
Our aim is to demonstrate the performance gain of our method on existing approaches when used as an aiding module to reduce the initial selection of the dictionary.
We also do not change the prior parameters when identifying the reduced models with the proposed VB inference steps.

To keep the discussion structured, we elaborate on each system and the respective implementation details in the related subsections. 
For all methods, we demonstrate the results of ordinary least squares solutions as given in \cite{edmd}, sequential thresholded least squares \cite{brunton2016discovering} solutions, Sparse Bayesian Learning (SBL) solutions \cite{biz} and finally the performance of the model identified with the proposed model.
The prior values of parameters used in the VB inference are presented in \ref{sec:apB}.

In our presentations, we use the labels I, II, III, IV for the least squares solution, sequential thresholded least squares solution \cite{brunton2016discovering}, SBL \cite{biz} and the proposed method, respectively.
For method II, we set the threshold parameter $\lambda = 0.05$ for the first system, $\lambda = 0.005$ for the second system and $\lambda = 0.015$ for the last system.

The first two systems investigated here are subject to artificially added measurement noise, the specifications of which are given in the respective sections.
The same level of noise is added to all the measured states independently.
This artificial measurement noise helps to examine our method's robustness against data corruption by evaluating the performance in changing signal to noise ratio (SNR) levels.
For these two systems, we consider $40$ Monte Carlo (MC) runs for each noise level and display the mean of the MC runs with solid lines.
The shaded areas represent the $95\%$ confidence interval for the corresponding color.
The last experiment includes real data, already corrupt with measurement noise and process noise.
Therefore, we do not add artificial noise and neither do we undertake a statistical study.
For all systems, we use the measured states and squared exponential kernels in the dictionary as per \cite{khosravi}. 
The centers of the kernel sections are set as the cluster centers of the clean states for a given cluster count.
For the last system, we also add squared exponentials with periodic exponents to the dictionary. 
For the exponent coefficients and frequencies, we set varying values.
All four methods are fitted using training only column normalization. 
Coefficients are mapped back to raw observable coordinates for prediction.
When a regressor column is divided by its training norm \(s_i\), its coefficient is multiplied by \(s_i\). 
The coefficient initialization variance and Gamma precision rate are therefore multiplied by \(s_i^2\), preserving their original raw coordinate definitions.

Our first performance metric is the one step prediction normalized mean squared error (NMSE) given by:
\begin{equation}
    \text{NMSE}_x = \frac{||x_\text{true} - x_\text{predicted}||^2}{||x_\text{true} - \text{mean}(x_\text{true})||^2},
\end{equation}
where the true states are not corrupted by artificial measurement noise.
 We also report the long term prediction NMSE performances of all methods.
Lastly, we investigate the condition numbers $\kappa$ of the autonomous block of the identified $\mathbf{A}^T$ in \eqref{eq:KF_blocks}.   
For the proposed method, this corresponds to the condition number of the elementwise product of two separately constructed matrices.

\subsubsection{Lorenz Attractor}
We begin by studying the proposed method on the Lorenz attractor. 
The attractor is a famous chaotic system that is commonly used as a nonlinear system identification benchmark \cite{lorenz_kaynak}.
We use simulation data of the system sampled at 1 KHz for 6 seconds for training and use unseen data for testing. 
Before dictionary construction and artificial noise injection, we divide the third Lorenz coordinate by five to reduce differences in coordinate magnitudes. 
The same transformation is used throughout training and testing.
For this system, the full size dictionary includes $23$ observables. 
For brevity, we only present the results of the $y$ state of the system as described in \cite{brunton2016discovering}.

\begin{figure}[t]
    \centering
    \includegraphics{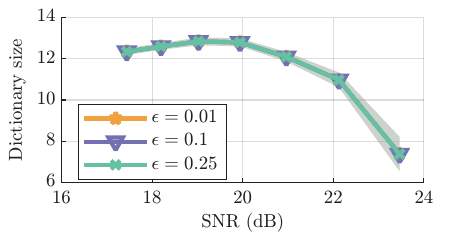}
    \caption{Reduced dictionary sizes of the Lorenz system with different $\epsilon$ across all measurement noise levels.}
    \label{fig:lorenz_ds}
\end{figure}

The average model size with respect to varying measurement noise strengths is given in Figure \ref{fig:lorenz_ds}.
It is seen that the Koopman model can be reduced to as low as 7 states depending on the noise level. 
Besides, the value of \(\epsilon\) is observed to have negligible effect.

\begin{figure}[t]
    \centering
    \includegraphics{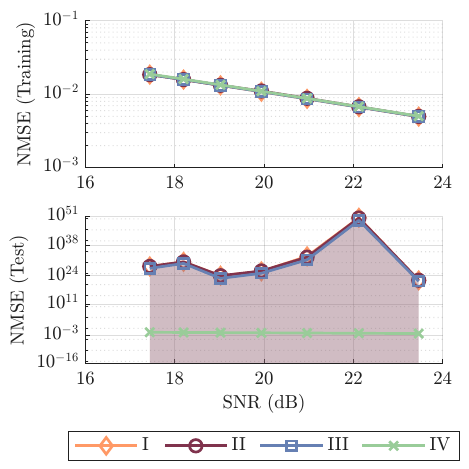}
    \caption{NMSE performance of full size models identified with different methods for the $y$ state of the Lorenz attractor in both training and test datasets.}
    \label{fig:lorenz_full}
\end{figure}

\begin{figure}[t]
    \centering
    \includegraphics{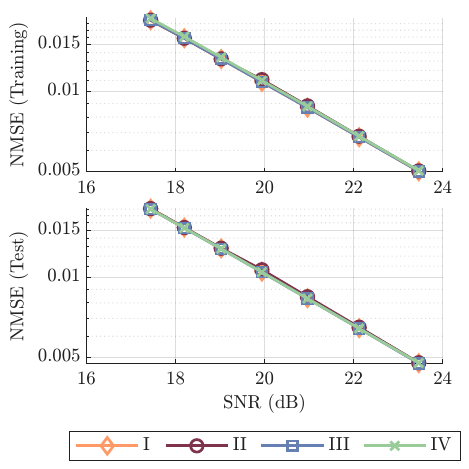}
    \caption{NMSE performance of reduced size models identified with different methods for the $y$ state of the Lorenz attractor in both training and test datasets.}
    \label{fig:lorenz_red}
\end{figure}

Figure \ref{fig:lorenz_full} demonstrates the NMSE performance of various methods using the full size dictionary.
Next, we determine the reduced dictionary $\Bar{\bphi}$ separately for each SNR level, then identify the reduced model parameters using all methods.
As a result of dictionary reduction, the reduced model size is different for each noise level. 
In Figure \ref{fig:lorenz_red}, we present the NMSE performance of all methods after the dictionary reduction has been applied as proposed with $\epsilon = 0.01$.
The large full dictionary errors are associated with narrow Gaussian observables with negligible training activation by inspection of the fitted models. 
Column normalization yields well conditioned fitting matrices but can produce large raw coordinate coefficients for these weakly supported features. 
Comparing Figure \ref{fig:lorenz_full} and Figure \ref{fig:lorenz_red}, dictionary reduction substantially lowers the test NMSE of Methods I–III, while Method IV largely maintains its prediction performance with fewer observables.
Dictionary selection removes many features with negligible training activation and substantially reduces the resulting errors. 


\begin{figure}[t]
    \centering
    \includegraphics{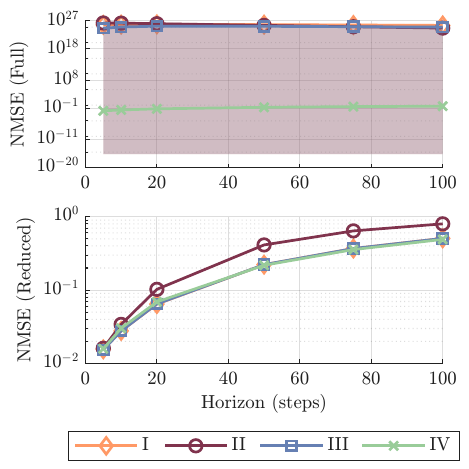}
    \caption{NMSE performances of the full and reduced models for the Lorenz attractor with respect to varying prediction horizons at a constant measurement noise level.}
    \label{fig:lorenz_longterm}
\end{figure}

We next investigate the long term prediction performance of all methods, both with full and reduced models.
We select a representative measurement noise level of near $20$ dB SNR and report the NMSE performances with respect to varying prediction horizons in Figure \ref{fig:lorenz_longterm}.
We start with the discrete prediction horizon length of $5$ and test different levels until $100$ using test trajectories.
We see that the reduced models maintain more reliable performance in the long prediction horizons compared to the full size models.
The difference is more prominent in methods I and II, which already present with very high one step NMSE performances in the full model.

\begin{figure}[t]
    \centering    
    \includegraphics[width=\linewidth]{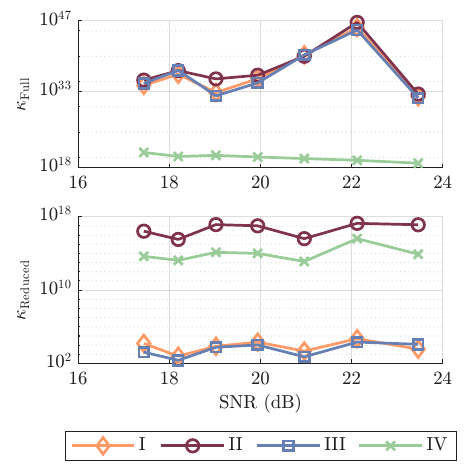}
    \caption{Average condition numbers of the full and reduced models of the Lorenz attractor}
    \label{fig:lorenz_kappa}
\end{figure}

\begin{figure}[t]
    \centering    
    \includegraphics[width=\linewidth]{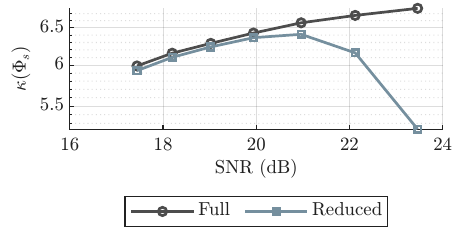}
    \caption{Condition numbers of the design matrix used in Lorenz experiment.}
    \label{fig:lorenz_phi}
\end{figure}

Finally, the mean condition numbers of the autonomous blocks of the identified $\hat{\bK}_F$ matrices and the design matrix \(\bPhi\) are reported in Figure \ref{fig:lorenz_kappa} and Figure \ref{fig:lorenz_phi}, respectively. 
These values characterize the singular value conditioning of the fitted transition matrices.
They do not directly establish the stability of forward predictions.
Prediction performance is evaluated separately in Figure \ref{fig:lorenz_longterm}, whereas sensitivity of coefficient estimation must be assessed using the regression design matrix \(\bPhi\).

The results in Figures \ref{fig:lorenz_longterm}, \ref{fig:lorenz_kappa} and \ref{fig:lorenz_phi} collectively suggest that the dictionary reduction is not only useful when the prediction performance is poor, but also when identified models are numerically sensitive despite the plausible performances in training data.
\subsubsection{Underactuated Unmanned Surface Vehicle (USV)}
We continue by implementing our method on a USV to demonstrate the performance on controlled systems.
We use the differential drive USV as characterized in \cite{simay} in a simulation environment. 
We generate the data at 10 Hz sampling frequency for 200 seconds for training.
In this study, only the velocity states and the control inputs are measured and included in the Koopman model. 
However, we add $1$ time step delayed states to the measured states and apply the observables to this combined state.
The full size dictionary includes $28$ observables of state and $2$ control inputs.
Dictionary reduction is not performed on the control inputs in any case.
To keep the exposition focused, we report results for only the surge velocity state; the remaining velocity components exhibit the same trend.

\begin{figure}[t]
    \centering
\includegraphics{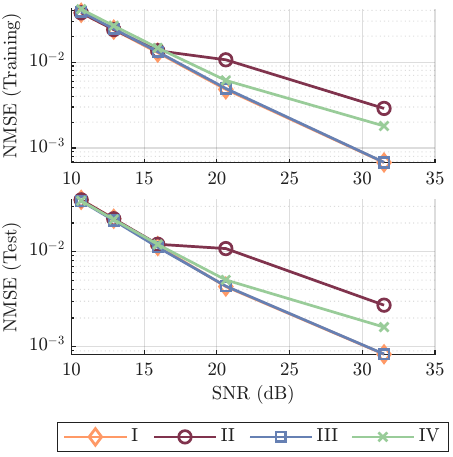}
    \caption{NMSE performance of full size models identified with different methods for the surge velocity state of USV in both training and test datasets.}
    \label{fig:heron_full}
\end{figure}

\begin{figure}[t]
    \centering
    \includegraphics{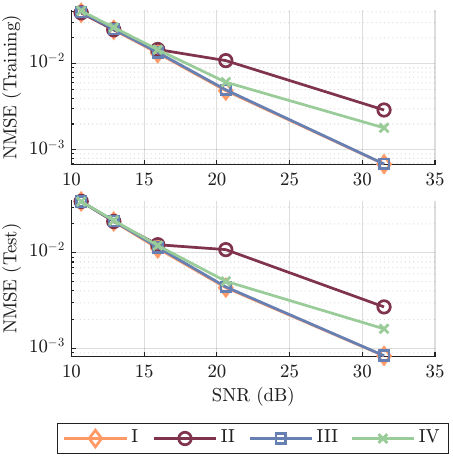}
    \caption{NMSE performance of reduced size models identified with different methods for the surge velocity state of USV in both training and test datasets.}
    \label{fig:heron_reduced}
\end{figure}

\begin{figure}[t]
    \centering
    \includegraphics{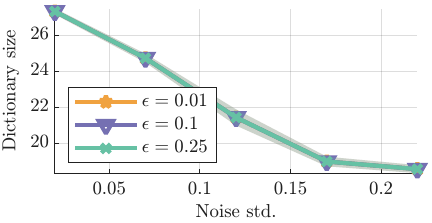}
    \caption{Reduced dictionary sizes of the USV with different $\epsilon$ across all measurement noise levels.}
    \label{fig:heron_ds}
\end{figure}

We provide the NMSE performance of all considered techniques using the full size dictionary in Figure \ref{fig:heron_full}.
This figure demonstrates that performances of all methods are comparable in lower SNR values while methods II and IV start to saturate learning in high SNR intervals.
This tendency is consistent with the strong sparsity bias in these methods, either set by the thresholding value or the parameter priors. 
We now provide NMSE performances of the methods with reduced dictionaries by setting $\epsilon = 0.25$ in Figure \ref{fig:heron_reduced}.
This is a more strict threshold when compared to the threshold we imposed for the Lorenz system. 
Comparing the full size model results in Figure \ref{fig:heron_full} with the reduced size model results in Figure \ref{fig:heron_reduced}, only minor differences are seen.
The performance of all the methods are largely unaltered by the reduced dictionary size.
The contribution of our method is therefore clear here, similar prediction performance is achieved through a smaller state space.
In particular, we present the dictionary sizes for three different $\epsilon$ thresholds for varying artificial measurement noise levels in Figure \ref{fig:heron_ds}.
As observed in Figure \ref{fig:heron_ds}, the retained dictionary sizes are similar across the tested thresholds. 
Moreover, the threshold level is observed to have negligible impact on the retained dictionary size in this experiment.  %

\begin{figure}[t]
    \centering
    \includegraphics{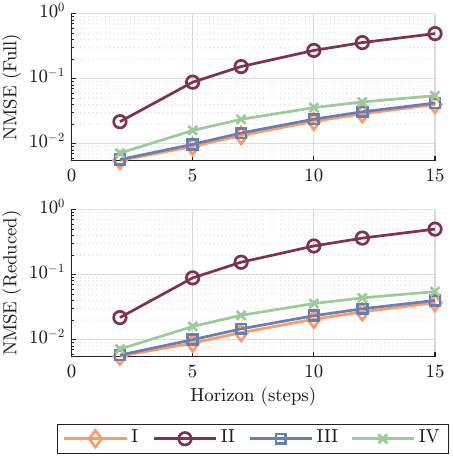}
    \caption{Long term prediction performances for the surge velocity state of USV on an unseen trajectory.}
    \label{fig:heron_longterm}
\end{figure}

Our next experiment is on the long term performances of the full and reduced size models for varying prediction horizons. 
We select discrete time prediction horizons ranging from $2$ to $15$ at the fixed SNR level of near 20 dB.
In Figure \ref{fig:heron_longterm}, we present the mean NMSE performances of all methods in previously unseen test cases.
Figure \ref{fig:heron_longterm} shows that the long term prediction performance is maintained despite the reduced dictionary.

\begin{figure}
    \centering    \includegraphics{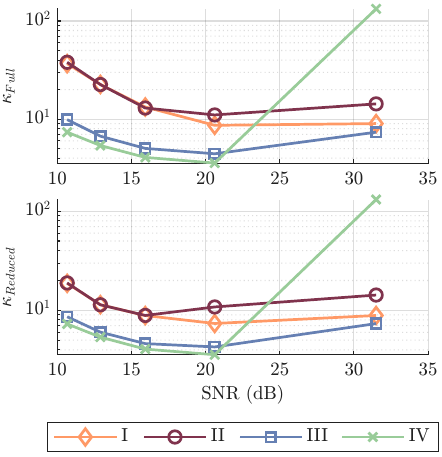}
    \caption{Condition numbers of the autonomous blocks of the full and reduced models of the USV model with respect to varying SNR levels.}
    \label{fig:heron_numerical}
\end{figure}

\begin{figure}
    \centering    \includegraphics{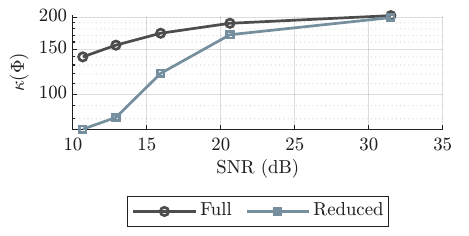}
    \caption{Condition numbers of the design matrix used in USV experiment}
    \label{fig:heron_phi}
\end{figure}

Finally, we compare the mean condition numbers of the full and
reduced USV autonomous transition blocks in
Figure~\ref{fig:heron_numerical} and the column-normalized
regression designs in Figure~\ref{fig:heron_phi}.
The connection between lifting design and numerical conditioning
has also been studied using reservoir based Koopman
representations \cite{gu2026koopman}.
Reduction does not uniformly improve the conditioning of the
autonomous transition blocks; its effect depends on the
identification method.
For the normalized regression designs, reduction lowers the mean condition numbers at all tested noise levels. 
The difference is small at the highest SNR, where dictionary reduction is also limited.
\subsubsection{Wiener-Hammerstein Process Noise System (WH)} 
Our final case study is the real data collected from a Wiener-Hammerstein system.
The system is a nonlinear operational amplifier-resistor network wrapped in between two LTI filters.
The filter at the output features a zero that is included in the input excitation frequency range.
The system is also subjected to both process and measurement noise injected via external sources, with process noise being the dominant disturbance.
The static nonlinearity caused by the diode networks is notoriously difficult to diagnose because the saturation causes information loss. 
Input signals with varying amplitudes and frequencies are used to capture the hybrid dynamics.
Another challenge in identifying this system is that the process noise enters before the nonlinear network, strengthening unpredictability \cite{WH}.

The full size dictionary consists of $46$ observables and one dimensional control input.
We apply our proposed dictionary reduction method by setting $\epsilon = 0.1$ and consequently reduce the dictionary to $15$ observables. 
We provide NMSE scores of full and reduced models in Table \ref{tab:WH}.
\begin{table}[t]
\centering
\caption{One step prediction NMSE scores of Koopman models for the Wiener-Hammerstein process noise system identified with different methods using the full and the reduced dictionaries.}
\begin{tabular}{@{}ccc@{}}
\toprule
\textbf{}        & \textbf{Training}                 & \textbf{Test}                     \\ \midrule
\textbf{Full}    & \textbf{}                         &                                   \\ \cmidrule(r){1-1}
\textbf{I}       & $0.003923$ & $0.004511$ \\
\textbf{II}      & $0.004027$ & $0.004552$  \\
\textbf{III}     & $0.003925$ & $0.00451$ \\
\textbf{IV}      & $0.004031$ & $0.004551$ \\ \midrule
\textbf{Reduced} & \textbf{}                         & \textbf{}                         \\ \cmidrule(r){1-1}
\textbf{I}       & $0.003928$ & $0.00451$ \\
\textbf{II}      & $0.00403$ & $0.00455$  \\
\textbf{III}     & $0.003928$ & $0.004509$ \\
\textbf{IV}      & $0.003935$ & $0.004516$  \\ 
\end{tabular}
\label{tab:WH}
\end{table}

In Table \ref{tab:WH}, we observe that dictionary reduction has not impacted the one step NMSE performance of the methods and all methods are seen to perform sufficiently in terms of one step NMSE.

However, for many systems applications, one step prediction performance can possibly be misleading \cite{pan2023tropaper}.
So, we examine the long term performances of the models with full and reduced dictionaries. 
We consider horizons ranging from $2$ to $50$ discrete time steps and test the methods on unseen trajectories.
Similarly to the one step NMSE results, we present the performances in Table \ref{tab:WH_long}.
\begin{table*}[]
\centering
\caption{Long term prediction NMSE performances of full and reduced models for the Wiener Hammerstein system.}
\begin{tabular}{@{}cccccc@{}}
\toprule
\textbf{Horizon length}         & \textbf{2}      & \textbf{5}      & \textbf{10}     & \textbf{20}      & \textbf{50}        \\ \midrule
\textbf{Full}    & \textbf{}     & & &                    &                                   \\ \cmidrule(r){1-1}
\textbf{I}      & $0.01803$ & $0.1089$ & $0.4007$ & $1.549$ & $3.786\times10^{4}$    \\
\textbf{II}     & $0.01807$ & $0.1088$ & $0.3929$ & $1.114$ & $1.686$    \\
\textbf{III}    & $0.01789$ & $0.1072$ & $0.383$ & $1.058$ & $1740$    \\
\textbf{IV}     & $0.0181$ & $0.1096$ & $0.3997$ & $1.154$ & $1.815$ \\
\midrule
\textbf{Reduced}    & \textbf{}     & & &                    &                                   \\ \cmidrule(r){1-1}
\textbf{I}   & $0.01789$ & $0.1074$ & $0.3845$ & $1.07$ & $1.549$    \\
\textbf{II}  & $0.01809$ & $0.1095$ & $0.3987$ & $1.148$ & $1.791$ \\
\textbf{III} & $0.01789$ & $0.1073$ & $0.3842$ & $1.068$ & $1.547$    \\
\textbf{IV}  & $0.01791$ & $0.1076$ & $0.3856$ & $1.076$ & $1.56$  \\\bottomrule
\end{tabular}
\label{tab:WH_long}
\end{table*}
Dictionary reduction substantially lowers the horizon-50 errors of Methods I and III and moderately improves Method IV. 
Method II shows a small deterioration. 
Although the objective function in the identification problem is not long term prediction performance, our results suggest that dictionary reduction and subsequent reidentification can improve long term performances.
Nevertheless, all reduced predictors have NMSE above unity at horizons 20 and 50, indicating limited long horizon accuracy relative to the reference mean benchmark. 
The result therefore supports compression and improved numerical behavior, while leaving long horizon prediction of this experiment unresolved.
These results show that dictionary reduction alone does not guarantee accurate long horizon prediction. 
Pruning can remove existing observables but cannot introduce missing observables or state histories.

Reducing the dictionary from 46 to 15 observables leaves one step prediction errors nearly unchanged, presented in Table \ref{tab:WH}.
It also substantially improves the conditioning of the regression design and autonomous transition blocks in Table \ref{tab:WH_cond}. 
Multi step benefits depend on the estimator.
Table \ref{tab:WH_long} shows methods I and III improve substantially at horizon 50, whereas Method II deteriorates slightly.
\begin{table}[t]
\centering
\caption{Condition numbers of the autonomous blocks of the identified $\hat{\bK}_F$ matrices of all methods and the condition number of the design matrix $\bPhi$ used in the methods with full and reduced dictionaries.}
\begin{tabular}{@{}ccc@{}}
\toprule
           & \textbf{Full} & \textbf{Reduced} \\ \midrule
\textbf{I}   & $2.802\times10^{33}$ & $4.455$                \\
\textbf{II}  & $8.132\times10^{31}$ & $4.179$                \\
\textbf{III} & $1.388\times10^{5}$ & $4.267$                \\
\textbf{IV}  & $1.55\times10^{23}$ & $1.64$\\
\midrule
    $\bPhi$ & $2.936\times10^{16}$ & $94.34$
\\ \bottomrule
\end{tabular}
\label{tab:WH_cond}
\end{table}

The results show that dictionary reduction improves numerical conditioning for all methods.
The design matrix $\bPhi$ is also seen to improve in terms of numerical conditioning. 

Overall, long term prediction performances and numerical properties show that dictionary reduction can be useful for many cases.
However, proposed hierarchical model is more useful as a dictionary reduction method than as a final Koopman model parameter estimation tool.
Once reduction is done, the reduced dictionary can be used with identification with less complex procedures as well.

\subsection{Preservation of indirect dependencies under dictionary reduction}
\label{subsec:structural_comparison}

In this section, we compare the proposed reduction method with pivoted QR \cite{businger1965leastSquares} and group lasso \cite{yuan2006groupLasso,obozinski2011support}.
The aim is to examine whether the selected observables retain
indirect dependencies and preserve the dynamics of the learned
VB model. 
We perform the comparison before refitting the
reduced model coefficients. 
Therefore, it measures preservation
of a common reference model rather than prediction accuracy against measured data.

Table~\ref{tab:structural_setup} gives the experimental settings.
We reuse the dictionaries, centers, VB fits, and graph thresholds
from the corresponding system experiments. Lorenz and USV use
40 measurement noise realizations each. 
WH uses one measured data
fit without additional artificial noise. 
All methods retain the original state coordinates and all input regressors.

\begin{table}[t]
    \centering
    \caption{Settings of the structural comparison.
    $L$ is the full number of observables, $M$ is the number of
    saved fits, $\sigma$ is the added measurement-noise standard
    deviation, $\epsilon$ is the graph threshold, and $h$ is the discrete
    propagation horizon. For WH, $\sigma=0$ means that no
    artificial noise is added.}
    \label{tab:structural_setup}
    \small
    \begin{tabular}{lrrrrr}
        \toprule
        System & $L$ & $M$ & $\sigma$ & $\epsilon$ & $h$ \\
        \midrule
        Lorenz & 23 & 40 & 0.9  & 0.01 & 100 \\
        USV    & 28 & 40 & 0.07 & 0.25 & 15  \\
        WH     & 46 & 1  & 0    & 0.10 & 50  \\
        \bottomrule
    \end{tabular}
\end{table}

The baseline methods select their dictionary sizes using an internal validation split within the original training data.

For each candidate support $\mathcal S$, where a support is the set of retained observable indices, we fit a predictor by ordinary least squares on the inner fitting block.
We select the support using the mean one step validation NMSE.
The references are noisy or measured outputs.
Each baseline selects its size independently.
Exact score ties favor fewer observables, followed by the earlier candidate in the selection path.

The dictionary is fixed before this internal split.
Its centers were obtained from the full original training block, including the part later used for validation.
The saved proposed model also uses the full original training block.
Thus, the validation data are held out from baseline coefficient fitting within a fixed dictionary. 
They are not held out from dictionary construction.

The proposed method retains the original state observables and all their ancestors in the thresholded graph.
This includes indirect predecessors and the intermediate observables along their paths.
The graph thresholds in Table~\ref{tab:structural_setup}
are kept fixed during the comparison.

We use column pivoted QR \cite{businger1965leastSquares}
to order the optional regressors. 
Regressors are normalized and projected off the mandatory state and input regressors before factorization. 
Each pivot prefix is fitted by least squares. We select the prefix with the lowest mean one step validation NMSE across the scored output channels.
We also use multivariate group lasso-based selection \cite{yuan2006groupLasso,obozinski2011support}, with mandatory states and inputs left unpenalized.
The penalized problems are solved using FISTA \cite{beck2009fast}.
Each converged support is refitted by unpenalized least squares and evaluated using the same validation criterion as QR. 
Nonconverged candidates are excluded.
For both methods, exact validation score ties favor
fewer observables.

After obtaining the three supports, we evaluate them against the same saved proposed thresholding operator.
Let $\widehat{\bK}_{\phi}\in\R^{L\times L}$ be its raw coordinate
posterior mean observable block, and let $\bG_{\phi}$ contain
the corresponding inclusion probabilities.
Let
$\bS=\diag(s_1,\ldots,s_L)$ contain the saved full-training
Euclidean norms of the observable columns.
The normalized, thresholded reference is defined by
\begin{equation}
    [\bK_{\epsilon}]_{ij}
    =
    \frac{s_i}{s_j}
    [\widehat{\bK}_{\phi}]_{ij}
    \mathbf{1}
    \!\left\{
        [\bG_{\phi}]_{ij}>\epsilon
    \right\}.
    \label{eq:structural_reference}
\end{equation}
After rescaling the coefficients in the matrix \(\widehat{\bK}_{\phi}\in\R^{L\times L}\) and thresholding by $\bG_{\phi}$, we obtain \(\bK_{\epsilon}\).
For each support, we retain the corresponding rows and columns of this reference matrix.
No coefficients are refitted in this structural test.
The least squares models used for validation are therefore separate from the operator used for evaluation.

We define two measures for this comparison.
First, let $\mathcal O$ contain the original state indices,
and let $d(i,\mathcal O)$ be the shortest directed path length
from observable $i$ to any index in $\mathcal O$.
The indirect-ancestor set is
\(
    \mathcal I
    =
    \left\{
        i:2\leq d(i,\mathcal O)<\infty
    \right\}.
\)
The indirect ancestor recall of a support $\mathcal S$ is
\begin{align}
    \rho_{\mathrm{ind}}(\mathcal S)
    =
    100\,
    \frac{
        |\mathcal S\cap\mathcal I|
    }{
        |\mathcal I|
    }.
    \label{eq:structural_recall}
\end{align}
We calculate recall for each fit and then take the mean.
Fits with no indirect ancestors are excluded from this measure because its denominator is zero.

The second measure is the relative multistep output map discrepancy.
Let $\bH\in\R^{L\times n}$ select the original $n$ state coordinates  and let
$\bP_{\mathcal S}\in\R^{L\times|\mathcal S|}$ contain the columns of the identity matrix indexed by $\mathcal S$.
The restricted operator and output selector are
\[
    \bK_{\mathcal S}
    =
    \bP_{\mathcal S}^{\top}
    \bK_{\epsilon}
    \bP_{\mathcal S},
    \qquad
    \bH_{\mathcal S}
    =
    \bP_{\mathcal S}^{\top}\bH.
\]
The full initial observable to output map at horizon $h$ is $\bK_{\epsilon}^{h}\bH$.
The restricted map, expressed in the full coordinates, is
\begin{align}
    \bR_h^{(\mathcal S)}
    =
    \bP_{\mathcal S}
    \bK_{\mathcal S}^{h}
    \bH_{\mathcal S}.
\end{align}
Multiplication by $\bP_{\mathcal S}$ inserts zero rows
at discarded observable indices.
The discrepancy is
\begin{equation}
    \delta_h(\mathcal S)
    =
    100\,
    \frac{
        \|\bK_{\epsilon}^{h}\bH
        -\bR_h^{(\mathcal S)}\|_F
    }{
        \|\bK_{\epsilon}^{h}\bH\|_F
    }.
    \label{eq:structural_discrepancy}
\end{equation}
All original states are evaluated for all systems.
This metric evaluates the zero-input observable to output map in the common normalized observable coordinates.
\begin{table*}[t]
    \centering
    \caption{Preservation of the thresholded VB model before
    refitting. Recall and discrepancy are percentages while the dictionary size refers to the mean retained dictionary size.
    Recall uses only fits containing indirect ancestors:
    three for Lorenz, 40 for USV, and one for WH.
    Discrepancy statistics use all saved fits.
    The proposed Lorenz result below $10^{-12}\%$ is
    consistent with numerical roundoff.}
    \label{tab:structural_results}
    \small
    \setlength{\tabcolsep}{4pt}
    \begin{tabular}{llrrrr}
        \toprule
        System & Selection
        & Mean $\rho_{\mathrm{ind}}$
        & Mean $\delta_h$
        & Max.\ $\delta_h$ 
        & Dictionary size\\
        \midrule
        Lorenz & Proposed
        & 100.0 & $<10^{-12}$ & $<10^{-12}$ & 12.8\\
        & Pivoted QR
        & 33.3 & 39.55 & 63.39 & 12.9\\
        & Group lasso
        & 33.3 & 7.09 & 54.08 & 14.8\\
        \midrule
        USV & Proposed
        & 100.0 & 0 & 0 & 24.73\\
        & Pivoted QR
        & 66.5 & 16.13 & 25.48 &22.62 \\
        & Group lasso
        & 55.6 & 25.99 & 54.92 & 20.45\\
        \midrule
        WH & Proposed
        & 100.0 & 0 & 0 & 15\\
        & Pivoted QR
        & 0.0 & 1.70 & 1.70 & 1\\
        & Group lasso
        & 0.0 & 1.70 & 1.70 & 1\\
        \bottomrule
    \end{tabular}
\end{table*}

Table~\ref{tab:structural_results} shows that the proposed
selection retains every inferred indirect ancestor and
preserves the thresholded output map up to roundoff.
This result follows from the selection rule.
Since the retained set contains all its predecessors,
\[
    (\bK_{\epsilon})_{\mathcal S^c,\mathcal S}=0.
\]
Discarded observables therefore cannot influence the
retained dynamics.
The numerical results confirm that the implementation
satisfies this property for the saved models.

USV provides the clearest example of the role of indirect dependencies.
QR and group lasso omit indirect ancestors and
give mean 15-step output-map discrepancies of $16.13\%$
and $25.99\%$, respectively.
The methods also retain different numbers of observables.
The mean sizes for the proposed method, QR, and group
lasso are $24.725$, $22.625$, and $20.45$ out of 28.
The proposed method therefore preserves the reference
dynamics with relatively limited compression in this case.

For Lorenz, indirect ancestors occur in only three of
the 40 fits. The output-map discrepancy can also result
from missing direct predecessors and changes in later
propagation. It cannot be explained only by indirect
dependencies.
Group lasso gives a lower mean discrepancy than QR,
but its maximum discrepancy is still substantial.
The mean retained sizes are $12.8$, $12.9$, and $14.825$
for the proposed method, QR, and group lasso, respectively.

For WH, both baselines select only the original scalar
state, while the proposed method retains 15 of the
46 observables.
Although the baselines omit all indirect ancestors,
the 50-step output-map discrepancy is only $1.70\%$.
Thus, missing graph dependencies do not always lead
to a large change in the evaluated output map.
This example shows that dependency preservation can require a larger dictionary than approximate preservation of the evaluated output map.

These results evaluate preservation of the learned thresholded model, rather than recovery of the true system dependencies.
The baselines are selected using one-step validation error, whereas the proposed method retains the dependencies required for exact output preservation before refitting.

\section{Conclusion}\label{sec:conclusion}
In this work, we have proposed a hierarchical probabilistic model for Koopman model reduction of discrete time dynamical systems.
We treat the regression of the time evolution of each observable independently from others.
The proposed framework combines probabilistic regressor target inclusion with graph ancestry search for output oriented dictionary reduction.
The performance of our method is evaluated in three representative systems. 
In the Lorenz attractor simulation we demonstrate that the reduced dictionary improves held-out prediction.
In the controlled USV case, we demonstrate that smaller dictionaries maintain the prediction performance,  with changes depending on the identification method.
Lastly in the real experimental data of the Wiener Hammerstein process, we demonstrate that the reduced dictionary can considerably improve numerical conditioning. 

On the other hand, our method requires prior tuning for many variables.
Since the initial dictionary is typically overparameterized, we recommend using sparsity inducing priors during the reduction stage to promote the elimination of redundant observables.
Prior information can also be incorporated by assigning stronger beliefs to observables that are known a priori to be essential for the model.
In particular, we recommend using sparsity inducing parameters when working with the initial dictionary.
After the dictionary has been reduced, however, the resulting model may benefit from less sparsity promoting priors, since excessive sparsity at this stage can lead to underfitting.
A further limitation of the proposed method is the use of a mean field approximation for posterior estimation, which may restrict the representation of dependencies among model parameters.   
Furthermore, our dictionary reduction method can only prune a given set of observables.
Consequently, the performance of the reduced models are still closely linked to the initially selected set of observables.

In our structural preservation comparisons, the proposed selection preserved the output maps of the proposed thresholded models up to numerical roundoff.
The comparison with QR and group lasso shows that selecting observables using one step validation error does not necessarily preserve the inferred multistep dynamics.

Although our method also provides an identification scheme for Koopman models without the reduction stage, we propose our method as a dictionary reduction tool to be used in conjunction with other identification schemes.
The selected dictionary can subsequently be used with any of the evaluated identification methods.

\appendix
\section{Derivations of VB inference updates}\label{sec:apA}
\subsection{Update of $q(\rho)$}
Following the VB inference rule \eqref{eq:posterior_star}, we write:
\begin{align}
    &\log q(\rho)    = \E_{q_{ \setminus \rho}}[\log p(\bt, \bbeta, \bgamma, \balpha, \bpi, \rho)],
\end{align}
where $p(\bt, \bbeta, \bgamma, \balpha, \bpi, \rho)$ is decomposed as in \eqref{eq:decomp_prior}.
The update of $\rho$ requires only $\rho$ dependent terms. 
As a result, we omit the terms where $\rho$ is not a variable:
\begin{align}
    &\log q(\rho) \peq \E_{q_{ \setminus \rho}}[\log p(\bt \mid \bbeta, \bgamma, \rho) + \log p(\rho)].
\end{align}
By \eqref{eq:wj_problem}, $p(\bt \mid \bbeta, \bgamma, \rho)$ is a Gaussian distribution and by \eqref{eq:rho_p}, $p(\rho)$ is a Gamma distribution.
Inserting the priors in the logarithm terms, we obtain:
\begin{align}
\begin{split}
     &\log q(\rho)  \peq \E_{q_{ \setminus \rho}} \bigg[ \sum_{k=1}^{m} \bigg( \log \frac{1}{\sqrt{2 \pi \rho^{-1}}} \\
    &- \frac{1}{2\rho^{-1}} (t_k - \phi(\bx[k])^T(\bgamma \odot \bbeta))^2 \bigg)
    + (a-1) \log \rho - b\rho \bigg],
\end{split}
\end{align}
\begin{align}
\begin{split}
  \peq \E_{q_{ \setminus \rho}} \bigg[ \frac{m}{2} \log \rho - \frac{\rho}{2} ||\bt - \Phi (\bgamma \odot \bbeta)||^2 + (a-1)\log \rho  \\  - b\rho \bigg].  
\end{split}
\end{align}
We model the posterior density $q(\rho)$ as a Gamma distribution. 
Therefore, we match the $\log \rho$ and $\rho$ terms on both sides of the equivalence:
\begin{align}
\begin{split}
    (\Bar{a} -1)\log \rho &= (\frac{m}{2} + a -1) \log \rho \\
    \implies \Bar{a} &= \frac{m}{2} + a.  
\end{split}
\end{align}
For \(\Bar{b}\), we define \(\bz = (\bgamma \odot \bbeta)\).
Then, the expectation is expanded as follows:
\begin{align}
\begin{split}
\Bar{b} \rho = \frac{\rho}{2}\E_{q_{ \setminus \rho}}[\sum_{k=1}^{m}(t_k - \phi(\bx[k])^T(\bgamma \odot \bbeta))^2] + b \rho\\
=  \sum_{k=1}^{m} \frac{\rho}{2}\E_{q_{ \setminus \rho}}[(t_k - \phi(\bx[k])^T(\bgamma \odot \bbeta))]^2 \\+ \sum_{k=1}^{m} \frac{\rho}{2} \text{Var}_{q_{ \setminus \rho}} [(t_k - \phi(\bx[k])^T(\bgamma \odot \bbeta))] + b \rho.   
\end{split}
\end{align}
The variance term is written as:
\begin{align}
\begin{split}
  \text{Var}_{q_{ \setminus \rho}}[(t_k& - \phi(\bx[k])^T(\bgamma \odot \bbeta))] \\
  &= \text{Var}_{q_{ \setminus \rho}} \bigg[t_k - \sum_{i=1}^{L+l} \bPhi_{ki} \gamma_i \beta_i \bigg]\\
  &= \sum_{i=1}^{L+l} \bPhi_{ki}^2 \big(\E_{q_{ \setminus \rho}} [\gamma_i^2 \beta_i^2] - \E_{q_{ \setminus \rho}} [\gamma_i \beta_i]^2 \big)\\
  &= \sum_{i=1}^{L+l} \bPhi_{ki}^2 \big( \hat{\gamma}_i (\mu_i^2 + \sigma_i^2) - \hat{\gamma}_i^2\mu_i^2 \big).    
\end{split}
\end{align}
Hence,
\begin{align}
\begin{split}
   \Bar{b} &= \frac{1}{2} \bigg[ ||\bt - \Phi(\hat{\bgamma} \odot \bmu)||^2\\ &+ \sum_{i=1}^{L+l} ||\bphi_{i}||^2 \big( \hat{\gamma}_i (\mu_i^2 + \sigma_i^2) - \hat{\gamma}_i^2\mu_i^2 \big) \bigg] +  b.    
\end{split}
\end{align}

\subsection{Update of $q(\alpha_i)$}

Starting by \eqref{eq:posterior_star} and the decomposition \eqref{eq:decomp_prior}, we write:
\begin{align}
    \log q(\alpha_i) = \E_{q_{ \setminus \alpha_i}} [\log p(\bt, \bbeta, \bgamma, \balpha, \bpi, \rho)],
\end{align}
where $\log p(\bt, \bbeta, \bgamma, \balpha, \bpi, \rho)$ is decomposed in \eqref{eq:decomp_prior}.
We discard the terms that do not depend on $\alpha_i$ since they can be normalized:
\begin{align}
\begin{split}
    \log q(\alpha_i) \peq\E_{q_{ \setminus \alpha_i}}[\log p(\bbeta \mid \balpha) + \log p(\balpha)],\\
    \peq \E_{q_{ \setminus \alpha_i}}[\log p(\beta_i \mid \alpha_i) + \log p(\alpha_i)],    
\end{split}
\end{align}
where the last step follows from the independence assumption.
Our priors indicate that $p(\beta_i \mid \alpha_i)$ is Gaussian as in \eqref{eq:beta_cp} and $p(\alpha_i)$ is Gamma as in \eqref{eq:alpha_p}.
Inserting the densities in the log terms and discarding constant terms in terms of $\alpha_i$, we get:
\begin{align}
\begin{split}
    &\log q(\alpha_i)\peq \\ &\E_{q_{ \setminus \alpha_i}} \bigg[ \log(\frac{1}{\sqrt{2 \pi \alpha_i^{-1}}}) - \frac{1}{2}\beta_i^2\alpha_i + (c_i -1)\log \alpha_i - d_i \alpha_i \bigg],   
\end{split}
\end{align}
\begin{align}
    &\peq \E_{q_{ \setminus \alpha_i}} \bigg[ \frac{1}{2}\log \alpha_i - \frac{\alpha_i}{2}\beta_i^2+(c_i-1)\log \alpha_i - d_i\alpha_i \bigg].
\end{align}
The expectation of $\frac{\alpha_i}{2}\beta_i^2$ is with respect to the posterior density $q(\beta_i)$, which is also Gaussian with mean $\mu_i$ and variance $\sigma_i^2$ and will be derived in the following steps. 
The expectation then becomes:
\begin{align}
     & \log q(\alpha_i) \peq \bigg(\frac{1}{2} + c_i -1\bigg)\log \alpha_i - \bigg(\frac{\mu_i^2 + \sigma_i^2}{2} + d_i \bigg) \alpha_i.
\end{align}
We model the posterior density $q(\alpha_i)$ as gamma distribution.
Accordingly,
\begin{align}
\begin{split}
(\Bar{c}_i -1)\log \alpha_i &= \bigg(\frac{1}{2} + c_i -1\bigg)\log \alpha_i\\
\implies \Bar{c}_i &= c_i + \frac{1}{2},   
\end{split}
\end{align}
\begin{align}
\begin{split}
\Bar{d}_i \alpha_i &= \bigg(\frac{\mu_i^2 + \sigma_i^2}{2} + d_i \bigg) \alpha_i, \\
    \implies \Bar{d}_i &= \bigg(\frac{\mu_i^2 + \sigma_i^2}{2} + d_i \bigg).
\end{split}
\end{align}
\subsection{Update of $q(\pi_i)$}
Once again, we start with the optimal posterior estimate \eqref{eq:posterior_star} and the prior decomposition \eqref{eq:decomp_prior}.
\begin{align}
    \log q(\pi_i) =\E_{q_{ \setminus \pi_i}} [\log p(\bt, \bbeta, \bgamma, \balpha, \bpi, \rho)].
\end{align}
Next, we omit all the terms that are independent of $\pi_i$:
\begin{align}
\begin{split}
    \log q(\pi_i) \peq\E_{q_{ \setminus \pi_i}} [\log p(\bgamma \mid \bpi) + \log p(\bpi)],\\
    \peq \E_{q_{ \setminus \pi_i}} [\log p(\gamma_i \mid \pi_i) + \log p(\pi_i)],    
\end{split}
\end{align}
where the last step is due to the independence assumption.
Our priors indicate that $p(\gamma_i \mid \pi_i)$ is a Bernoulli distribution and $p(\pi_i)$ is a Beta distribution.
We insert the corresponding distributions inside the log terms;
\begin{align}
\begin{split}
    \log q(\pi_i) \peq \E_{q_{\setminus \pi_i}}[\gamma_i\log \pi_i + (1-\gamma_i)\log (1-\pi_i) \\+ (e_i-1)\log \pi_i + (f_i-1)\log (1-\pi_i)]\\ 
\end{split}
\end{align}
\begin{align}
\peq \E_{q_{\setminus \pi_i}}[(\gamma_i + e_i - 1)\log \pi_i + (1-\gamma_i + f_i-1)\log(1-\pi_i)].
\end{align}
Taking the expectation, we have;
\begin{align}
\begin{split}
    &\log q(\pi_i) \peq \\ &(\hat{\gamma}_i + e_i -1)\log \pi_i + (1-\hat{\gamma}_i + f_i-1)\log(1-\pi_i).
\end{split}
\end{align}
We model the posterior density $q(\pi_i)$ as a Beta distribution.
Accordingly, we match the corresponding terms on both sides of the equivalence:
\begin{align}
\begin{split}
    (\Bar{e}_i - 1) \log \pi_i &= (\hat{\gamma}_i + e_i -1)\log \pi_i\\
    \implies \Bar{e}_i &= \hat{\gamma}_i + e_i,   
\end{split}
\end{align}
\begin{align}
\begin{split}
(\Bar{f}_i - 1) \log(1-\pi_i) &= (1-\hat{\gamma}_i + f_i-1)\log(1-\pi_i)\\
\implies \Bar{f}_i &=  (1-\hat{\gamma}_i + f_i)  .      
\end{split}
\end{align}

\subsection{Update of $q(\beta_i)$}
Similar to every other random variable update, we start with the optimal posterior rule \eqref{eq:posterior_star} and the prior decomposition \eqref{eq:decomp_prior}.
\begin{align}
    \log q(\beta_i) = \E_{q_{ \setminus \beta_i}}[\log p(\bt, \bbeta, \bgamma, \balpha, \bpi, \rho)]
\end{align}
The terms without $\beta_i$ dependence are discarded as they can be normalized out of the equations:
\begin{align}
    &\log q(\beta_i) \peq \E_{q_{ \setminus \beta_i}}[\log p(\bt \mid \bbeta, \bgamma, \rho) + \log p(\bbeta \mid \balpha)].
\end{align}
According to the priors, $p(\bt \mid \bbeta, \bgamma, \rho)$ and $p(\bbeta \mid \balpha)$ are Gaussian distributions.
Specifically, we can further discard terms in $p(\bbeta \mid \balpha)$ and only consider $p(\beta_i \mid \alpha_i)$ as the priors are independent.
Similarly, in order to leave the $\beta_i$ related terms alone, we manipulate $\log p(\bt \mid \bbeta, \bgamma, \rho)$ term:
\begin{align}
\begin{split}
&\log p(\bt \mid \bbeta, \bgamma, \rho) \peq m \log \frac{1}{\sqrt{2\pi \rho^{-1}}} \\&- \frac{\rho}{2} \sum_{k=1}^m \bigg(t_k - \sum_{ \substack{j=1 \\ j\neq i}}^{L+l}\gamma_j\beta_j\phi_j(\bx[k]) - \gamma_i\beta_i\phi_i(\bx[k]) \bigg)  ^2,
\end{split}
\end{align}
\begin{align}
\begin{split}
& \peq  \sum_{k=1}^{m} \frac{-\rho}{2} \bigg( \gamma_i \beta_i^2 \phi_i^2(\bx[k]) \\& - 2\gamma_i\beta_i\phi_i(\bx[k])\bigg(t_k - \sum_{ \substack{j=1 \\ j\neq i}}^{L+l}\gamma_j\beta_j\phi_j(\bx[k]) \bigg) \bigg). 
\end{split}
    \label{eq:pt_dissection}
\end{align}
By defining the residual $r_{ik}$ as in \eqref{eq:r_ik}, we replace $\log p(\bt \mid \bbeta, \bgamma, \rho)$ inside $\log q(\beta_i)$:
\begin{align}
\begin{split}
\log q(\beta_i)
\peq \E_{q_{ \setminus \beta_i}}\bigg[ \sum_{k=1}^{m} \frac{-\rho}{2} \bigg( \gamma_i \beta_i^2 \phi_i^2(\bx[k]) \\ - 2\gamma_i\beta_i\phi_i(\bx[k])r_{ik} \bigg)
        -\frac{\alpha_i}{2}\beta_i^2 \bigg],      
\end{split}
\end{align}
\begin{align}
\begin{split}
    \peq \E_{q_{ \setminus \beta_i}} \bigg[ \beta_i^2 \bigg( \sum_{k=1}^{m} \frac{-\rho}{2}\gamma_i \phi_i^2(\bx[k]) -\frac{\alpha_i}{2} \bigg) + \\ 
     \beta_i \bigg( \sum_{k=1}^{m} \rho \gamma_i\phi_i(\bx[k]) r_{ik} \bigg) \bigg] .      
\end{split}
\end{align}
We model the posterior $q(\beta_i)$ as a Gaussian distribution.
Substituting $\log q(\beta_i)$ and matching both sides of the equivalence, we have:
\begin{align}
    \Bar{\alpha}_i := \Bar{\sigma}_i^{-2} = \sum_{k=1}^{m} \hat{\rho}\hat{\gamma}_i \phi_i^2(\bx[k]) + \hat{\alpha}_i,\\
    \bar{\mu}_i = \frac{\sum_{k=1}^{m} \hat{\rho}\hat{\gamma}_i\phi_i(\bx[k]) \hat{r}_{ik}}{\sum_{k=1}^{m} \hat{\rho}\hat{\gamma}_i \phi_i^2(\bx[k]) + \hat{\alpha}_i}.
\end{align}
\subsection{Update of $q(\gamma_i)$}
We once again start from \eqref{eq:posterior_star} and the prior decomposition \eqref{eq:decomp_prior}.
Accordingly,
\begin{align}
    \log q(\gamma_i) =\E_{q_{ \setminus \gamma_i}}[\log p(\bt, \bbeta, \bgamma, \balpha, \bpi, \rho)]
\end{align}
We discard the terms that do not include $\gamma_i$ as a variable and get:
\begin{align}
\begin{split}
     \log q(\gamma_i) \peq \E_{q_{ \setminus \gamma_i}}[\log p(\bt \mid \bbeta, \bgamma, \rho) + \log p(\bgamma \mid \bpi)], 
\end{split}
\end{align}
\begin{align}
\begin{split}
    \peq \E_{q_{ \setminus \gamma_i}}[\log p(\bt \mid \bbeta, \bgamma, \rho) + \log p(\gamma_i \mid \pi_i)],    
\end{split}
\end{align}
where the last step is due to independence of the priors.
We substitute the values of the prior densities $p(\bt \mid \bbeta, \bgamma, \rho)$ and $p(\gamma_i \mid \pi_i)$ which are Gaussian and Bernoulli distributions, respectively.
In doing so, we dissect $\log(p(\bt \mid \bbeta, \bgamma, \rho)$ as in \eqref{eq:pt_dissection} such that $\gamma_i$ terms are left alone: 
\begin{align}
\begin{split}
    \log q(\gamma_i) \peq \E_{q_{ \setminus \gamma_i}} \bigg[ \frac{-\rho}{2} \sum_{k=1}^{m} (r_{ik} - \gamma_i \beta_i \phi_i(\bx[k]))^2  +\\ \gamma_i \log(\pi_i) + (1-\gamma_i)\log(1-\pi_i) \bigg]   . 
\end{split}
\end{align}
By defining $\bm{r}_i$ as in \eqref{eq:br} and $\bphi_i$ as
\begin{align}
    \bphi_i = \begin{bmatrix}
                  \phi_i(\bx[1]) & \dots & \phi_i(\bx[m])
              \end{bmatrix}^T
              \label{eq:bphi_i}
\end{align}
for all $i=1, \dots, L+l,$ we simplify the notation and obtain
\begin{align}
\begin{split}
     \log q(\gamma_i) \peq \E_{q_{ \setminus \gamma_i}} \bigg[ \frac{-\rho}{2} (-2 \gamma_i \beta_i \bm{\phi}_i^T\bm{r}_i + \gamma_i \beta_i^2 ||\bm{\phi}_i||^2)                    \\
    + \gamma_i \log(\pi_i) + (1-\gamma_i)\log(1-\pi_i) \bigg].   
\end{split}
\end{align}
Taking the expectation,
\begin{align}
\begin{split}
     \log q(\gamma_i) \peq \frac{-\hat{\rho}}{2}(-2 \gamma_i \mu_i \bm{\phi}_i^T\hat{\bm{r}}_i \\+ \gamma_i (\mu_i^2 + \sigma_i^2) ||\bm{\phi}_i||^2)
    + \gamma_i(\psi(\bar{e}_i) - \psi(\bar{f}_i)),
    \label{eq:log_gamma_genel_appendix}   
\end{split}
\end{align}
where
\begin{align}
    \begin{split}
        \E_{\pi_i}[\log(\pi_i)]    & = \psi(\bar{e_i}) - \psi(\bar{e}_i+\bar{f}_i), \\
        \E_{\pi_i}[\log(1- \pi_i)] & = \psi(\bar{f_i}) - \psi(\bar{e}_i+\bar{f}_i)
    \end{split}
    \label{eq:digamma_appendix}
\end{align}
and $\psi(\cdot)$ is the Digamma function \cite{spouge1994computation}.
Continuing from \eqref{eq:log_gamma_genel_appendix}, we group the terms together and simply avoid the constant terms since they can be normalized:
\begin{align}
    \log q(\gamma_i) & \peq \gamma_i \eta_i,
    \label{eq:log_gamma_eta}
\end{align}
where
\begin{align}
    \eta_i = \bigg( \hat{\rho}\mu_i\bm{\phi}_i^T\hat{\bm{r}}_i - \frac{\hat{\rho}}{2}(\mu_i^2 + \sigma_i^2) ||\bm{\phi}_i||^2 + \psi(\bar{e}_i) - \psi(\bar{f}_i)\bigg).
    \label{eq:eta_defn_appendix}
\end{align}
Since we model the posterior distribution of $\gamma_i$ as a Bernoulli distribution, $\gamma_i$ can take values in $\{0,1\}$.
Then we may write
\begin{align}
    \begin{split}
        q(\gamma_i)     & \propto e^{\gamma_i \eta_i},       \\
q(\gamma_i=1) &= \frac{e^{\eta_i}}{1+e^{\eta_i}},\\
q(\gamma_i=0) &= \frac{1}{1+e^{\eta_i}}.
    \end{split}
    \label{eq:gamma_weigths_appendix}
\end{align}
There can only be two such weights in \eqref{eq:gamma_weigths_appendix}, so we can normalize the weights to get $q(\gamma_i)$.
As a result,
\begin{align}
    \bar{\pi}_i = \frac{e^{\eta_i}}{1 + e^{\eta_i}} = \frac{1}{1+ e^{-\eta_i}} = \sigma(\eta_i),
\end{align}
where $\sigma(\cdot)$ is the sigmoid function.

\section{Parameter values in the experiments}\label{sec:apB}
Tables\ref{tab:lorenz_priors}-\ref{tab:WH_priors} report the hyperprior parameters and the initial
values of the variational distributions. 
The parameters
$a$, $b$, $c$, $d$, $e$, and $f$ define the hyperprior
distributions introduced in Section \ref{sec:method}.
The quantity $\sigma_0^2$ is the initial variance of
$q(\beta_i)$, and $\bar{\pi}_0$ is the initial inclusion
probability of $q(\gamma_i)$. 
All initial coefficient means are zero. 
The coefficient-related values are specified before column normalization; the coefficient variance and the Gamma rate parameter $d$ are transformed as described in Section \ref{sec:results}.
\subsection{Lorenz attractor}
\begin{table}[t]
\centering
\setlength{\tabcolsep}{2.8pt}
\begin{tabular}{@{}ccccccccc@{}}
\toprule
\textbf{Parameter} & $a$ & $b$ & $c$ & $d$ & $e$ & $f$ & $\sigma^2$ & $\pi$ \\ \midrule
\textbf{Value}     & 1.01       & 0.1        & $10^{-3}$       & $10^{-4}$      & 0.5        & 20         & 10                     & 0.5         \\ \bottomrule
\end{tabular}
\caption{Hyperprior parameters and variational initialization
for the Lorenz experiment.}
\label{tab:lorenz_priors}
\end{table}
Table \ref{tab:lorenz_priors} provides the values of the parameters used in the VB inference for the Lorenz attractor.

\subsection{USV}
Table \ref{tab:heron_priors} provides the values of the parameters used in the VB inference for the USV system.
\begin{table}[t]
\centering
\setlength{\tabcolsep}{2.8pt}
\begin{tabular}{@{}ccccccccc@{}}
\toprule
\textbf{Parameter} & $a$ & $b$ & $c$ & $d$ & $e$ & $f$ & $\sigma^2$ & $\pi$ \\ \midrule
\textbf{Value}     & 1.01       & 0.1        & $10^{-4}$       & $10^{-4}$      & 0.5        & 20         & 10                     & 0.5         \\ \bottomrule
\end{tabular}
\caption{Hyperprior parameters and variational initialization
for the USV experiment.}
\label{tab:heron_priors}
\end{table}

\subsection{Wiener Hammerstein system}
Table \ref{tab:WH_priors} provides the values of the parameters used in the VB inference for the Wiener Hammerstein process noise system.
\begin{table}[t]
\centering
\setlength{\tabcolsep}{2.8pt}
\begin{tabular}{@{}ccccccccc@{}}
\toprule
\textbf{Parameter} & $a$ & $b$ & $c$ & $d$ & $e$ & $f$ & $\sigma^2$ & $\pi$ \\ \midrule
\textbf{Value}     & 1       & 0.1        & $10^{-3}$       & $10^{-3}$      & 1        & 1         & 10                     & 0.5         \\ \bottomrule
\end{tabular}
\caption{Hyperprior parameters and variational initialization
for the Wiener--Hammerstein experiment.}
\label{tab:WH_priors}
\end{table}

\bibliography{refs}

@ARTICLE{khosravi,
  author={Khosravi, Mohammad},
  title={Representer Theorem for Learning {Koopman} Operators}, 
  journal={IEEE Transactions on Automatic Control},
  volume={68},
  number={5},
  pages={2995--3010},
  year={2023},
  publisher={IEEE},
  doi={10.1109/TAC.2023.3242325}}

@article{brunton2016discovering,
  title={Discovering governing equations from data by sparse identification of nonlinear dynamical systems},
  author={Brunton, Steven L and Proctor, Joshua L and Kutz, J Nathan},
  journal={Proceedings of the national academy of sciences},
  volume={113},
  number={15},
  pages={3932--3937},
  year={2016},
  publisher={National Academy of Sciences}
}

@article{abraham2017model,
  title={Model-based control using {Koopman} operators},
  author={Abraham, Ian and De La Torre, Gerardo and Murphey, Todd D.},
  journal={arXiv preprint arXiv:1709.01568},
  howpublished={arXiv preprint},
  year={2017}
}

@article{shi2024koopmansurvey,
  title={Koopman operators in robot learning},
  author={Shi, Lu and Haseli, Masih and Mamakoukas, Giorgos and Bruder, Daniel and Abraham, Ian and Murphey, Todd and Cort{\'e}s, Jorge and Karydis, Konstantinos},
  journal={IEEE Transactions on Robotics},
  year={2026},
  publisher={IEEE}
}

@article{lusch2018deep,
  title={Deep learning for universal linear embeddings of nonlinear dynamics},
  author={Lusch, Bethany and Kutz, J. Nathan and Brunton, Steven L.},
  journal={Nature Communications},
  volume={9},
  number={1},
  pages={4950},
  year={2018},
  publisher={Nature Publishing Group UK London}
}

@book{mauroy2020koopman,
  title={{Koopman} operator in systems and control},
  author={Mauroy, Alexandre and Susuki, Y. and Mezic, Igor},
  volume={484},
  year={2020},
  publisher={Springer}
}

@article{koopman1931hamiltonian,
  title={Hamiltonian systems and transformation in {Hilbert} space},
  author={Koopman, Bernard O},
  journal={Proceedings of the National Academy of Sciences},
  volume={17},
  number={5},
  pages={315--318},
  year={1931}
}

@article{dmd,
  title={Dynamic mode decomposition of numerical and experimental data},
  author={Schmid, Peter J},
  journal={Journal of fluid mechanics},
  volume={656},
  pages={5--28},
  year={2010},
  publisher={Cambridge University Press}
}

@article{edmd,
  title={A data--driven approximation of the {Koopman} operator: Extending dynamic mode decomposition},
  author={Williams, Matthew O and Kevrekidis, Ioannis G and Rowley, Clarence W},
  journal={Journal of Nonlinear Science},
  volume={25},
  number={6},
  pages={1307--1346},
  year={2015},
  publisher={Springer}
}

@article{koopmandictionary,
  title={Extended dynamic mode decomposition with dictionary learning: A data-driven adaptive spectral decomposition of the {Koopman} operator},
  author={Li, Qianxiao and Dietrich, Felix and Bollt, Erik M and Kevrekidis, Ioannis G},
  journal={Chaos: An Interdisciplinary Journal of Nonlinear Science},
  volume={27},
  number={10},
  year={2017},
  publisher={AIP Publishing}
}

@article{koopmanwithcontrol,
  title={Generalizing {Koopman} theory to allow for inputs and control},
  author={Proctor, Joshua L and Brunton, Steven L and Kutz, J Nathan},
  journal={SIAM Journal on Applied Dynamical Systems},
  volume={17},
  number={1},
  pages={909--930},
  year={2018},
  publisher={SIAM}
}

@article{em,
  title={Maximum likelihood from incomplete data via the {EM} algorithm},
  author={Dempster, Arthur P and Laird, Nan M and Rubin, Donald B},
  journal={Journal of the royal statistical society: series B (methodological)},
  volume={39},
  number={1},
  pages={1--22},
  year={1977},
  publisher={Wiley Online Library}
}

@book{bishop2006pattern,
  title={Pattern recognition and machine learning},
  author={Bishop, Christopher M. and Nasrabadi, Nasser M.},
  volume={4},
  number={4},
  year={2006},
  publisher={Springer}
}

@ARTICLE{simay,
  author={Atasoy, Simay and Karagöz, Osman Kaan and Ankarali, Mustafa Mert},
  journal={IEEE Access}, 
  title={Trajectory-Free Motion Planning of an Unmanned Surface Vehicle Based on {MPC} and Sparse Neighborhood Graph}, 
  year={2023},
  volume={11},
  number={},
  pages={47690-47700},
  doi={10.1109/ACCESS.2023.3275433}}

@ARTICLE{vb,
title={The variational approximation for {Bayesian} inference},
author={Tzikas, Dimitris G and Likas, Aristidis C and Galatsanos, Nikolaos P},
journal={IEEE Signal Processing Magazine},
volume={25},
number={6},
pages={131--146},
year={2008},
publisher={IEEE},
doi={10.1109/MSP.2008.929620}}

@article {lorenz_kaynak,
  title={Deterministic nonperiodic flow},
  author={Edward Norton Lorenz},
  journal={Journal of the Atmospheric Sciences},
  year={1963},
  volume={20},
  pages={130-141},
  url={https://api.semanticscholar.org/CorpusID:15359559}
}

@inproceedings{biz,
  title={Sparse {Bayesian} Learning for {Koopman} Based System Identification},
  author={Özcan, Selin Ezgi and Ankaralı, Mustafa Mert},
  booktitle={2025 11th International Conference on Control, Decision and Information Technologies (CoDIT)},
  year={2025},
  volume={1},
  pages={1888--1893},
  year={2025},
  organization={IEEE}
}

@article{spikeslab,
  author  = {H. Ishwaran and J. S. Rao},
  title   = {Spike and Slab Variable Selection: Frequentist and {Bayesian} Strategies},
  journal = {The Annals of Statistics},
  volume  = {33},
  number  = {2},
  pages   = {730--773},
  year    = {2005},
  month   = apr,
  doi     = {10.1214/009053604000001147}
}

@misc{WH,
  author       = {Schoukens, Maarten and No{\"e}l, Jean-Philippe},
  title        = {Wiener-Hammerstein Benchmark with Process Noise},
  year         = {2020},
  publisher    = {4TU.ResearchData},
  version      = {2},
  howpublished = {Dataset},
  doi          = {10.4121/12952124.v2},
  url          = {https://doi.org/10.4121/12952124.v2}
}

@ARTICLE{pan2023tropaper,
  author={Pan, Jie and Li, Dongyue and Wang, Jian and Zhang, Pengfei and Shao, Jinyan and Yu, Junzhi},
  journal={IEEE Transactions on Robotics}, 
  title={Autogeneration of Mission-Oriented Robot Controllers Using {Bayesian}-Based {Koopman} Operator}, 
  year={2024},
  volume={40},
  number={},
  pages={903-918},
  doi={10.1109/TRO.2023.3344033}}

@article{spouge1994computation,
  title     = {Computation of the gamma, digamma, and trigamma functions},
  author    = {Spouge, John L},
  journal   = {SIAM Journal on Numerical Analysis},
  volume    = {31},
  number    = {3},
  pages     = {931--944},
  year      = {1994},
  publisher = {SIAM}
}

@misc{TPE,
      title={Tree-Structured {Parzen} Estimator: Understanding Its Algorithm Components and Their Roles for Better Empirical Performance}, 
      author={Shuhei Watanabe},
      year={2026},
      eprint={2304.11127},
      archivePrefix={arXiv},
      primaryClass={cs.LG},
      url={https://arxiv.org/abs/2304.11127}, 
}

@book{saad2003iterative,
  title={Iterative methods for sparse linear systems},
  author={Saad, Yousef},
  year={2003},
  publisher={SIAM}
}

@article{nuutila1994finding,
  title     = {On finding the strongly connected components in a directed graph},
  author    = {Nuutila, Esko and Soisalon-Soininen, Eljas},
  journal   = {Information processing letters},
  volume    = {49},
  number    = {1},
  pages     = {9--14},
  year      = {1994},
  publisher = {Elsevier}
}

@article{Vu_2024,
   title={R-VGAL: a sequential variational {Bayes} algorithm for generalised linear mixed models},
   volume={34},
   ISSN={1573-1375},
   url={http://dx.doi.org/10.1007/s11222-024-10422-8},
   DOI={10.1007/s11222-024-10422-8},
   number={3},
   journal={Statistics and Computing},
   publisher={Springer Science and Business Media LLC},
   author={Vu, Bao Anh and Gunawan, David and Zammit-Mangion, Andrew},
   year={2024},
   month=apr }

@article{Nayek_2021,
   title={On spike-and-slab priors for {Bayesian} equation discovery of nonlinear dynamical systems via sparse linear regression},
   volume={161},
   ISSN={0888-3270},
   url={http://dx.doi.org/10.1016/j.ymssp.2021.107986},
   DOI={10.1016/j.ymssp.2021.107986},
   journal={Mechanical Systems and Signal Processing},
   publisher={Elsevier BV},
   author={Nayek, R. and Fuentes, R. and Worden, K. and Cross, E.J.},
   year={2021},
   month=dec, pages={107986} }

@misc{thompson2022computationallimitsdeeplearning,
      title={The Computational Limits of Deep Learning}, 
      author={Neil C. Thompson and Kristjan Greenewald and Keeheon Lee and Gabriel F. Manso},
      year={2022},
      eprint={2007.05558},
      archivePrefix={arXiv},
      primaryClass={cs.LG},
      url={https://arxiv.org/abs/2007.05558}, 
}

@article{jordan1999introduction,
  title={An introduction to variational methods for graphical models},
  author={Jordan, Michael I and Ghahramani, Zoubin and Jaakkola, Tommi S and Saul, Lawrence K},
  journal={Machine learning},
  volume={37},
  number={2},
  pages={183--233},
  year={1999},
  publisher={Springer}
}

@article{graphdecomp,
      title={Sparse decompositions of nonlinear dynamical systems and applications to moment-sum-of-squares relaxations}, 
      author={Corbinian Schlosser and Milan Korda},
      journal={Mathematics of Control, Signals, and Systems},
      pages={1--31},
      year={2026},
      DOI={10.1007/s00498-026-00443-1},
      publisher={Springer}
}

@article{tarjan1972depth,
  title={Depth-first search and linear graph algorithms},
  author={Tarjan, Robert},
  journal={SIAM journal on computing},
  volume={1},
  number={2},
  pages={146--160},
  year={1972},
  publisher={SIAM}
}

@article{gu2026koopman,
  title={Koopman Identification of Nonlinear Systems via Reservoir Liftings},
  author={Gu, Weibin and Yang, Chen and Shi, Lu},
  journal={IEEE Control Systems Letters},
  year={2026},
  publisher={IEEE}
}

@inproceedings{
miao2026learning,
title={Learning {Koopman} Representations with Controllability Guarantees},
author={Keyan Miao and Han Wang and Xuda Ding and Konstantinos Gatsis and Andreas Krause and Antonis Papachristodoulou},
booktitle={The Fourteenth International Conference on Learning Representations},
year={2026},
url={https://openreview.net/forum?id=jITPFROpWN}
}

@article{pan2021sparsity,
  title={Sparsity-promoting algorithms for the discovery of informative Koopman-invariant subspaces},
  author={Pan, Shaowu and Arnold-Medabalimi, Nicholas and Duraisamy, Karthik},
  journal={Journal of Fluid Mechanics},
  volume={917},
  pages={A18},
  year={2021},
  publisher={Cambridge University Press}
}

@article{schlosser2022sparsity,
  title={Sparsity Structures for Koopman and Perron--Frobenius Operators},
  author={Schlosser, Corbinian and Korda, Milan},
  journal={SIAM Journal on Applied Dynamical Systems},
  volume={21},
  number={3},
  pages={2187--2214},
  year={2022},
  publisher={SIAM}
}

@article{businger1965leastSquares,
  title={Linear least squares solutions by Householder transformations},
  author={Businger, Peter and Golub, Gene H},
  journal={Numerische Mathematik},
  volume={7},
  number={3},
  pages={269--276},
  year={1965},
  publisher={Springer}
}

@article{yuan2006grouplasso,
  title={Model selection and estimation in regression with grouped variables},
  author={Yuan, Ming and Lin, Yi},
  journal={Journal of the Royal Statistical Society Series B: Statistical Methodology},
  volume={68},
  number={1},
  pages={49--67},
  year={2006},
  publisher={Oxford University Press}
}

@article{obozinski2011support,
  title={Support union recovery in high-dimensional multivariate regression},
  author={Obozinski, Guillaume and Wainwright, Martin J and Jordan, Michael I},
  year={2011},
  journal = {The Annals of Statistics},
  volume  = {39},
  number  = {1},
  pages   = {1--47},
  doi     = {10.1214/09-AOS776}
}

@article{beck2009fast,
  title={A fast iterative shrinkage-thresholding algorithm for linear inverse problems},
  author={Beck, Amir and Teboulle, Marc},
  journal={SIAM journal on imaging sciences},
  volume={2},
  number={1},
  pages={183--202},
  year={2009},
  publisher={SIAM}
}

\end{document}